\documentclass[aps,prx,superscriptaddress,amsfonts,amsmath,amssymb,showpacs,floatfix,reprint,longbibliography,footinbib,]{revtex4-2}

\usepackage[english]{babel}
\usepackage{times}
\usepackage{url}
\usepackage{bm}
\usepackage{graphicx}
\usepackage{amsmath}
\usepackage{amstext}
\usepackage{amssymb}
\usepackage{amsfonts}
\usepackage{amsbsy}
\usepackage{verbatim}
\usepackage{xcolor}
\definecolor{dkgreen}{rgb}{0, 0.5, 0}
\definecolor{midnightblue}{rgb}{0.39, 0.58, 0.93}
\definecolor{darkblue}{HTML}{004D6B}
\definecolor{darkred}{HTML}{8c1515}
\definecolor{darkgreen}{HTML}{006400}
\usepackage{multirow}
\usepackage{floatrow}
\usepackage{float}
\usepackage{tikz}
\usepackage{gensymb}
\usepackage{textcomp}
\usepackage{enumitem}
\usepackage[version=3]{mhchem}
\usepackage{afterpage}
\usepackage{pbox}
\usepackage{makecell}
\usepackage{dsfont}
\usepackage{stackengine,wasysym,scalerel}
\usepackage{latexsym}
\usepackage{cancel}
\usepackage{array, multirow, bigdelim, makecell, booktabs}
\usepackage[misc]{ifsym}
\usepackage[percent]{overpic}
\usepackage[normalem]{ulem}
\usepackage{mathtools}
\usepackage{pict2e}
\usepackage{bbm}
\usepackage{setspace}
\usepackage{dcolumn}
\usepackage{placeins}
\usepackage{mathrsfs}
\usepackage{bbold}
\usepackage{amsthm}
\usepackage{epsfig}
\usepackage{array}
\usepackage{color}
\usepackage{physics}
\usepackage{soul}
\usepackage{silence}
\usepackage{comment}
\usepackage[colorlinks=true, urlcolor=midnightblue, linkcolor=midnightblue, citecolor=midnightblue]{hyperref}
\usepackage[capitalise]{cleveref}

\newcolumntype{P}[1]{>{\centering\arraybackslash}p{#1}}

\makeatletter
\edef\textFontName{\fontname\csname
	\f@encoding/\f@family/\f@series/\f@shape/\f@size\endcsname}

\makeatother

\def\bra#1{\left\langle#1\right|}
\def\ket#1{\left|#1\right\rangle}

\mathchardef\sOmega="710A
\mathchardef\sGamma="7100
\mathchardef\sDelta="7101

\DeclareMathOperator*{\sumint}{
	\mathchoice
	{\ooalign{$\displaystyle\sum$\cr\hidewidth$\displaystyle\int$\hidewidth\cr}}
	{\ooalign{\raisebox{.14\height}{\scalebox{.7}{$\textstyle\sum$}}\cr\hidewidth$\textstyle\int$\hidewidth\cr}}
	{\ooalign{\raisebox{.2\height}{\scalebox{.6}{$\scriptstyle\sum$}}\cr$\scriptstyle\int$\cr}}
	{\ooalign{\raisebox{.2\height}{\scalebox{.6}{$\scriptstyle\sum$}}\cr$\scriptstyle\int$\cr}}
}

\newcommand{\cre}{{\dag}}
\newcommand{\ann}{{\vphantom{\dag}}}
\newcommand{\noprime}{{\vphantom{\prime}}}

\newcommand{\I}{\text{i}}

\newcommand{\xvortex}[8]{\Gamma_{xx\,i_{#7}i_{#8}}^{\Lambda\,#1'#2'|#3^\noprime#4^\noprime}(#5,#6|#3^\noprime,#4^\noprime)}
\newcommand{\yvortex}[8]{\Gamma_{yy\,i_{#7}i_{#8}}^{\Lambda\,#1'#2'|#3^\noprime#4^\noprime}(#5,#6|#3^\noprime,#4^\noprime)}
\newcommand{\zvortex}[8]{\Gamma_{zz\,i_{#7}i_{#8}}^{\Lambda\,#1'#2'|#3^\noprime#4^\noprime}(#5,#6|#3^\noprime,#4^\noprime)}
\newcommand{\dvortex}[8]{\Gamma_{d\,i_{#7}i_{#8}}^{\Lambda\,#1'#2'|#3^\noprime#4^\noprime}(#5,#6|#3^\noprime,#4^\noprime)}
\newcommand{\dird}[4]{\delta(\omega_{#3}+\omega_{#4}-\omega_{#1}-\omega_{#2})}

\makeatletter
\def\convertto#1#2{\strip@pt\dimexpr #2*65536/\number\dimexpr 1#1}
\makeatother
\newcommand{\measurewidths}{{%
		\color{red}
		:::
		TEXT \convertto{in}{\the\textwidth} in
		:::
		COLUMN \convertto{in}{\the\columnwidth} in
		:::
		HEIGHT(ex) \convertto{pt}{1ex} pt
		:::
		WIDTH(em) \convertto{pt}{1em} pt
		:::
}}

\renewcommand{\arraystretch}{1.35}

\AtBeginDocument{\RenewCommandCopy\qty\SI}
\begin{document}
	
	
	\title{Keldysh pseudo-fermion functional renormalization group for quantum magnetism}
	
	\author{Janik Potten}
	\email{Corresponding author: janik.potten@uni-wuerzburg.de}
	\affiliation{Institute for Theoretical Physics and Astrophysics, University of W\"urzburg, D-97074 W\"urzburg, Germany}
	
	\author{Yasir Iqbal}
	\affiliation{Department of Physics and Quantum Centre of Excellence for Diamond and Emergent Materials (QuCenDiEM), Indian Institute of Technology Madras, Chennai 600036, India}
	
	\author{Ronny Thomale}
	\affiliation{Institute for Theoretical Physics and Astrophysics, University of W\"urzburg, D-97074 W\"urzburg, Germany}
	\affiliation{Department of Physics and Quantum Centre of Excellence for Diamond and Emergent Materials (QuCenDiEM), Indian Institute of Technology Madras, Chennai 600036, India}
	
	\author{Tobias Müller}
	\affiliation{Institute for Theoretical Physics and Astrophysics, University of W\"urzburg, D-97074 W\"urzburg, Germany}
	\affiliation{Department of Physics, University of Zurich, Winterthurerstrasse 190, 8057 Zurich, Switzerland}
	
	\date{\today}

	
	\begin{abstract}
		The functional renormalization group (FRG) approach for spin models relying on a pseudo-fermionic description has proven to be a powerful technique in simulating ground state properties of strongly frustrated magnetic lattices. A drawback of the FRG framework is that it is formulated in the imaginary-time Matsubara formalism and thus only able to access static correlations, a limitation shared with most other many-body approaches. A description of the dynamical properties of magnetic systems is the key to bridging the gap between theory and neutron scattering spectra. We take the decisive step of expanding the scope of pseudo-fermion FRG to the Keldysh formalism, which, while originally developed to address non-equilibrium phenomena, enables a direct calculation of the equilibrium dynamical spin structure factors on generic lattices in arbitrary dimension. We identify the principal features characterizing the low-energy spectra of exemplary zero-, one- and two-dimensional spin-$1/2$ Heisenberg models as well as the Kitaev honeycomb model while identifying current limitations of the method that have to be improved upon.
	\end{abstract}

	\maketitle

	
	\section{Introduction} The field of frustrated magnetism is currently poised with the arrival of materials based on novel two- and three-dimensional lattice geometries promising to host a wealth of exotic quantum phases~\cite{Chamorro-2021}. These span complex spin textures where the dipole moments form helices~\cite{Binz-2006}, skyrmions~\cite{Muhlbauer-2009,Nagaosa-2013}, hedgehogs~\cite{Fujishiro-2019}, vortex crystals or commensurate noncoplanar structures~\cite{Messio-2011,Gembe-2023,Gembe-2024}, as well as unconventional magnetic phases characterized by mutipole order parameters~\cite{Santini-2009,Kuramoto-2009} such as quadrupolar (spin-nematic state)~\cite{Andreev-1984}, octupolar~\cite{Sibille-2020}, or even higher-rank orders~\cite{Ikeda-2012,Paddison-2015}. These phases feature nontrivial spin wave excitation spectra with characteristic fingerprints which allow for their identification based on inelastic neutron scattering profiles~\cite{Smerald-2013,Kato-2021}. The most intriguing scenario occurs when the system displays no magnetic order (either dipolar or multipolar) down to arbitrarily low temperatures, no sign of translation symmetry breaking, and no sharp magnon dispersion as it supports fractionalized spinon excitations, i.e., those quasiparticles formed in a {\it quantum spin liquid} (QSL) state~\cite{Savary-2017,Broholm-2020}. Being generally defined by zero-temperature measurements, it is these very properties of QSLs rendering them notoriously difficult to describe theoretically, and to detect experimentally in comparison with other quantum fluids studied in condensed matter systems. Since, in most crystals, the strictly elastic limit is dominated by crystal defects and imperfections, one rather studies the data in the weakly inelastic domain above the disorder scale, which is still expected to share many features of the static signature in the idealized clean case. Thus, the most indicative, measurable quantity at hand is the dynamical spin structure factor obtained through inelastic neutron scattering.
	
	Theoretically, there is neither a satisfactory microscopic description nor a fully adequate mean field theory available for QSLs. As much as major progress has been made within parton approaches and spin liquid ground states, at the mean-field level, can now be adequately classified at the model level by structures such as their gauge group~\cite{Wen-2002}, there is still a disconnect between trial states generated from mean field theories and the microscopic Hamiltonian's ground state. From mean field theory, we further know that distinct types of spin liquids can look alike in terms of {\it static} correlations in finite size simulations yet might be significantly different in terms of their {\it dynamic} correlation profile. The problem here is that one can only generate static, but not dynamic, spin correlators from ground state wave functions. While recent attempts~\cite{BeccaSpinChain,VMCSquareLattice} have been made at computing dynamical spin structure factors beyond mean-field, they are limited in scope by the class of excitations of the matter fields that can be constructed while gauge-field excitations, which are known to be crucial for QSLs, are not {\it a priori} accounted for. The current effort of the theory community in frustrated magnetism thus focuses on finding ways to access dynamical spin correlations in higher dimensional quantum paramagnets. In doing so, in addition to acquiring a better identification tool for spin liquids states, one also attempts to address the fundamental question of how $S=1$ magnon excitations in an ordered magnet dissociate into fractional $S=1/2$ spinon excitation as one approaches the paramagnetic regime across a quantum critical point. A convincing theoretical analysis has so far only been accomplished in spin chains~\cite{Pereira-2006}. In two spatial dimensions, where analytically exact limits are less available than for spin chains, progress has been recently reported from density matrix renormalization group (DMRG) studies~\cite{Sherman-2023,Drescher-2023}. Similarly, variational Monte Carlo methods can in principle be generalized to track energy expectation values of excited trial states. In both cases, however, frustrated magnets in 3D, which is the regime where current new material candidates for spin liquids are predominantely identified~\cite{Gonzalez-2024,Chillal-2020,Plumb-2019}, pose a nearly unsurmountable challenge either due to entanglement growth (DMRG) or due to increases in the number of wave function parameters (Monte Carlo).
	
	In order to calculate dynamical spin correlations for 2D and, in particular, 3D quantum paramagnets, here we develop a Keldysh formulation of the pseudofermion functional renormalization group (PFFRG)~\cite{PFFRGReview}. Our starting point is the Matsubara formulation of the PFFRG~\cite{ReutherFirstFRG}, as it was refined and optimized in recent years. While the Keldysh form is suited to calculate non-equilibrium phenomena as well~\cite{Kamenev,JakobsSIAM,Nonequilibrium,Sieberer-2016,Frassdorf-2017,Muenchen1,Muenchen2}, our predominant motivation lies in avoiding analytic continuation from the imaginary to the real frequency axis as we attempt to access the dynamical reponse functions, which have so far been inaccessible within the PFFRG (except for particular cases in which a Pad\'e approximation was applicable~\cite{AnalyticalContinuation}).
	
	This paper is organized as follows. After a short introduction of Abrikosov pseudo-fermions (\cref{sec:pseudofermions}) and the main building blocks of the Keldysh formalism (\cref{sec:Keldysh}), we formulate the Keldysh pseudo-fermion functional renormalization group in \cref{sec:Kpffrg}. There, we also detail its classical large-$S$ limit (\cref{sec:larges}), as well as performing detailed symmetry analysis of the relevant vertex functions (\cref{sec:symmetry}), the definition of relevant observables (\cref{sec:observables}) and the numerical implementation (\cref{sec:numerics}). We benchmark the method in \cref{sec:results} for the zero-dimensional Heisenberg dimer (\cref{sec:dimer}), the one-dimensional spin chain (\cref{sec:spinchain}), as well as the two-dimensional square lattice (\cref{sec:square}) and Kitaev-Honeycomb (\cref{sec:kitaev}) models. Finally, we conclude in \cref{sec:conclusion} that our Keldysh implementation of the pseudofermion FRG promises to be a key advancement in rendering the dynamical fingerprint of a frustrated magnet accessible within the scope of the functional renormalization group.
	
	\section{Abrikosov Pseudo Fermions}\label{sec:pseudofermions}
	In this paper we consider a family of possibly anisotropic spin Hamiltonians 
	\begin{align}\label{eq:ham}
		H=\!\sum_{\substack{i,j\\\mu,\nu}}J^{\mu\nu}_{ij}S^\mu_iS^\nu_j,
	\end{align}
	which includes i.e., the Heisenberg model with $J^{\mu\nu}_{ij}\equiv J_{ij}\delta_{\mu\nu}$ and the Kitaev model. It can be derived as a strong coupling limit from the Hubbard model at half-filling and is well suited to describe localized interacting electron spins. However, due to the canonical commutation relations of the spin operators, which are neither fermionic nor bosonic, and the resulting inapplicability of Wick's theorem, standard many-body techniques are not straightforwardly applicable to \eqref{eq:ham} \cite{PFFRGReview}. Especially, a rigorous formulation of the flow equations using the spin operators proves to be cumbersome~\cite{kriegExactRenormalization2019} and they have so far only be solved in limiting cases~\cite{tarasevychRichMan2018,gollSpinFunctional2019,gollZeromagnonSound2020,tarasevychDissipativeSpin2021,tarasevychCriticalSpin2022,ruckriegelSpinFunctional2022,ruckriegel2024phase}, while a numerical implementation is still at large.
	
	One way of circumventing this obstacle is to introduce an auxiliary fermion representation \cite{Abrikosov} of the spin operators
	\begin{align}\label{eq:decomposition}
		S^\mu_i=\sum_{\beta,\beta'}\frac{1}{2} f^\cre_{i\beta}\sigma^\mu_{\beta\beta'}f^\ann_{i\beta'}.  
	\end{align}
    where $\beta$ describes the spin projection in the $z$-direction and can either be $\uparrow$ or $\downarrow$.
	With this reformulation, the Hamiltonian takes the form
	
	\begin{align}\label{eq:HamPF}
		H_\text{PF}=\!\sum_{\substack{i,j\\\mu,\nu}}\sum_{\beta^\noprime_1,\beta^\prime_1}\sum_{\beta^\noprime_2,\beta^\prime_2}\frac{J^{\mu\nu}_{ij}}{4}\sigma^\mu_{\beta^\noprime_1\beta^\prime _1}\sigma^\nu_{\beta^\noprime_2\beta^\prime_2}f^\cre_{i\beta^\noprime_1}f^\cre_{j\beta^\noprime_2}f^\ann_{j\beta^\prime_2}f^\ann_{i\beta^\prime_1},
	\end{align}
	
	To render the mapping in \eqref{eq:decomposition} exact, one has to introduce the constraint
	\begin{align}\label{eq:PFConstraint}
		\sum_{\beta}f^\cre_{i\beta}f^\ann_{i\beta}=1,
	\end{align}
	on the particle number since the fermion decomposition doubles the Hilbert space and allows also for empty or doubly-occupied sites, which do not map to a physical spin state. For this reason, these fermions are often referred to as pseudo-fermions. 
	There is a technique originally proposed by V. Popov and S. Fedotov \cite{Popov,Popov2} to fulfill this constraint exactly using an imaginary chemical potential but its treatment is out of the scope of this paper [see Refs.~\cite{BenediktPopov, DissReuther, kiselev2, kiselevreview}].
	Further, there is also a pseudo-Majorana approach  for the Heisenberg model \cite{PseudoMajorana1,PseudoMajorana2}, which is able to circumvent this constraint with the decomposed Hilbert space only being physical. This will be a future path for our technique and is also not further discussed here. 
	
	In the following we assume that the constraint is not fulfilled exactly but rather on average. This will be investigated in detail for the anti-ferromagnetic (AFM) dimer in Sec.~\ref{sec:dimer}.
	
	\section{Keldysh Formalism}\label{sec:Keldysh}
	The Keldysh formalism was originally proposed as a framework to describe non-equilibrium phenomena \cite{Keldysh}. Here, we employ it to introduce real frequencies into our formalism, diverging from the conventional use of imaginary Matsubara frequencies, while still maintaining equilibrium conditions, as in \cite{kiselev1}.
	We will only provide a brief overview of the main ideas and refer to the literature \cite{Kamenev} for a detailed introduction.
	
\begin{figure}
		\centering
		\includegraphics[width=0.98\textwidth]{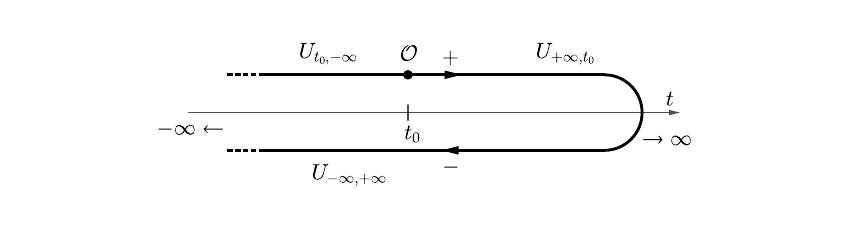}
		\caption{Keldysh contour with an operator $\mathcal{O}$ inserted at $t=t_0$. The respective time propagation operators are also given.}
		\label{fig:KeldyshContour}
	\end{figure}

	Non-equilibrium dynamics potentially evolve the ground state beyond acquiring a dynamical phase, adding another dimension of complexity to the calculation of expectation values.
	To illustrate, we consider a generic, time-dependent Hamiltonian $H(t)$. The system is then described by the time-evolution operator 
	\begin{align}
		U_{t,t'}=\mathrm{T}\exp\left(-\I\int\limits_{t'}^t H(\tau)\mathrm{d}{\tau}\right),\label{eq:timeevoop}
	\end{align}
	where $\mathrm{T}$ denotes time ordering. Assuming the system is initially in the state described by the density matrix $\rho(-\infty)$, at time $t$, it is given by
	\begin{align}
		\rho(t)=U_{t,-\infty}\rho(-\infty)U_{t,-\infty}^\dagger.
	\end{align}
	Using this, the expectation value of an observable $\mathcal{O}$ at a given time $t_0$ can be expressed as:
	\begin{align}\nonumber
		\expval{\mathcal{O}(t_0)}&=\tr[\mathcal{O}\rho(t_0)]/\tr[\rho(t_0)]\\
		&=\tr[U_{-\infty,t_0}\mathcal{O}U_{t_0,-\infty}\rho(-\infty)]/\tr[\rho(t_0)]\\
		&=\tr[U_{-\infty,\infty}U_{\infty,t_0}\mathcal{O}U_{t_0,-\infty}\rho(-\infty)]/\tr[\rho(-\infty)]\nonumber.
	\end{align}
	In the last step we introduced $U_{t_0,\infty}U_{\infty,t_0}$ as a representation of unity which formally closes the time contour of the Keldysh formalism at $t=-\infty$. This time contour, depicted in Fig.~\ref{fig:KeldyshContour}, consists of two branches, which are the forward-propagating (denoted by $+$) and backward-propagating ($-$) one. 
	
	When calculating expectation values, the operators thus have be to contour ordered (denoted by $\mathcal{C}$), because placing an operator on the forward propagating branch can have a different effect than on the backward propagating branch.
	This also necessitates four different Green's functions 
	\begin{align}
		\begin{pmatrix}
			G^{++} & G^{+-}\\
			G^{-+} & G^{--}
		\end{pmatrix}\equiv
		\begin{pmatrix}
			G^{T} & G^{<}\\
			G^{>} & G^{\tilde{T}}
		\end{pmatrix},
	\end{align}
	which are given by $\mathrm{i}G^{\eta\eta'} = \langle\mathcal{C}f^\eta f^{\eta'}\rangle$ with $\eta^{(\prime)}=\pm$. The latter notation consisting of time ordered ($G^T$), anti-time ordered $(G^{\tilde{T}})$, lesser ($G^{<}$) and greater ($G^{>}$) Green's functions is typically found in the literature. 
	
	The formal closure of the contour introduces an ambiguity in the formalism, as at this point no contour index can be assigned. This leads to a redundancy in the Green's function, which can be removed by performing the basis transformation  
	
	\begin{align}\label{eq:Keldysh rotation}
		A=\frac{1}{\sqrt{2}}\begin{pmatrix}
			1 & -1\\
			1 & 1
		\end{pmatrix},
	\end{align}
	which results in 
	\begin{align}
		A
		\begin{pmatrix}
			G^{++} & G^{+-}\\
			G^{-+} & G^{--}
		\end{pmatrix}
		A^{-1}=
		\begin{pmatrix}
			0 & G^{\text{av}}\\
			G^{\text{ret}} & G^{\text{K}}
		\end{pmatrix},
	\end{align}
	where $G^\text{ret}$ and $G^\text{av}$ are the typical retarded and advanced Green's functions and $G^\text{K}$ is an additional Keldysh Green's function. In equilibrium, the latter is fully determined by the fluctuation dissipation theorem
	\begin{align}\label{eq:FDT}
		G^\text{K}(\omega)=(1-2n_\text{F}(\omega))(G^\text{ret}(\omega)-G^\text{av}(\omega)).
	\end{align}
	The independent Green's functions for a free-fermion with $H_0=\epsilon_0 f^\cre f^\ann$ in Fourier space are given by
	\begin{align}\label{eq:freeFermionPropagators}
		G^\text{ret/av}_0(\omega)=\lim_{\eta\to0}\frac{1}{\omega-\epsilon_0\pm\I\eta}.
	\end{align}
	For both formulas we used the Fourier transformation given by
	\begin{align}\label{eq:FourierToOmega}
		G(\omega)&=\quad\,\,\,\int\mathrm{e}^{\mathrm{i}\omega t}G(t)\,\mathrm{d}t,\\
		G(t)&=\frac{1}{2\pi}\int\mathrm{e}^{-\mathrm{i}\omega t}G(\omega)\,\mathrm{d}\omega.\label{eq:FourierToTime}
	\end{align}
	
	\section{Pseudo-fermion Functional Renormalization Group}\label{sec:Kpffrg}
	In this section, we bring the concepts of auxiliary fermions and the Keldysh formalism together with the functional renormalization group (FRG). The general framework of RG is based on Wilson's idea for quantum many-particle systems~\cite{WilsonRG} which formulates an effective low energy theory by integrating out all high energy degrees of freedom. This is done by artificially introducing a cutoff parameter $\Lambda$ into our theory $G_0\rightarrow G_0^\Lambda$ which fulfills the properties
	\begin{align}\label{eq:conditions}
		G_{0}^{\Lambda=\infty}=0,\quad G_{0}^{\Lambda=0}=G_0.
	\end{align}
	This is not necessarily the only starting condition for $\Lambda=\infty$ in the FRG formalism since there exist other i.e. from a correlated initial state for the fermionic Hubbard-model \cite{DMF2RG}. However, shutting off all particle dynamics is very practical in our case, arriving at the conditions above. Since, our Hamiltonian is quartic in terms of the pseudo-fermions we have a free theory with no kinetic (quadratic) term. We can thus use the free particle Green's functions~\eqref{eq:freeFermionPropagators} with $\epsilon_0=0$. For the cutoff, there are in principle different choices~\cite{TemperatureFlow}, because the conditions~\eqref{eq:conditions} are relatively weak but it has been shown that a cutoff using the convergence parameter for advanced and retarded Green's functions preserves the most symmetries of the system -- especially causality -- and thus should be used~\cite{DissJakobs,JakobsSymmetries}. The bare propagators thus read
	\begin{align}
		G^{\Lambda\,\text{ret/av}}_0(\omega)=\frac{1}{\omega\pm\I\Lambda},
	\end{align}
	while the Keldysh component is still given via the fluctuation-dissipation theorem~(\ref{eq:FDT}).
	Employing this cutoff we can formulate differential equations which allow us to calculate the low energy properties of the system by successively integrating in degrees of freedom. 
	For the FRG we formulate these flow equations directly in terms of the one-particle irreducible vertex functions.
	
	Compared with the more standard Matsubara formulation~\cite{PFFRGReview}, the derivation of the Keldysh flow equations only deviates at two points. First, there is an imaginary unit present in the time evolution operator Eq.~\eqref{eq:timeevoop} and second the definitions of the Green's functions likewise feature another imaginary $\mathrm{i}$. These exactly cancel each other~\cite{Karrasch,KarraschDiplom}, leaving the actual flow equations form-invariant between the two formulations.
	For the one- and two-particle vertex functions $\gamma_1$ and $\gamma_2$ these are given by
	
	\begin{align}
		&\frac{\mathrm{d}}{\mathrm{d}\Lambda}\gamma_1^\Lambda(1'|1)=\sumint_{22'}\gamma_2^\Lambda(1',2'|1,2)S^\Lambda(2|2'),
	\end{align}
	and
	\begin{align}
		&\frac{\mathrm{d}}{\mathrm{d}\Lambda}\gamma_2^\Lambda(1,'2'|1,2)=\sumint_{33'}\gamma_3^\Lambda(1',2',3'|1,2,3)S^\Lambda_{33'}\nonumber\\	\label{eq:flowEq}
		&\quad+\sumint_{33'}\sumint_{44'}\big[\gamma_2^\Lambda(1',2'|3,4)\gamma^\Lambda_2(3',4'|1,2)\\
		&\quad-\gamma_2^\Lambda(1',4'|1,3)\gamma_2^\Lambda(3',2'|4,2)-(3^{(\prime)}\leftrightarrow4^{(\prime)})\nonumber\\[12pt]
		&\quad+\gamma_2^\Lambda(2',4'|1,3)\gamma_2^\Lambda(3',1'|4,2)+(3^{(\prime)}\leftrightarrow4^{(\prime)})\big]G^\Lambda_{33'}S^\Lambda_{44'}\nonumber.
	\end{align}
	Here, we have used multi-indices $k$ which consist of all relevant quantum numbers for the problem at hand. $S$ denotes the so-called single-scale propagator
	\begin{equation}
		S^\Lambda = G^\Lambda \left[\partial_\Lambda (G_0^\Lambda)^{-1}\right] G^\Lambda.
	\end{equation}
	As illustrated in Eq.~\eqref{eq:flowEq}, the $n$-particle vertex always couples to the $n+1$-particle vertex in the flow. Therefore, to arrive at a closed set of differential equations, this hierarchical structure has to be broken, which we do by neglecting all $n>2$-particle contributions, setting $\gamma_{n>2}=0$. As usual for the PFFRG this is remedied partially by using the Katanin substitution
	\begin{equation}
		S^\Lambda \to \frac{\mathrm{d}}{\mathrm{d}\Lambda} G^\Lambda,
	\end{equation}
	which was shown to fulfill the Ward identities up to one order higher \cite{Katanin} for certain diagrams.  
	The effect of the truncation will be discussed in the following sections.
	
	Furthermore, using the tree expansion one can relate the single particle vertex function to the self-energy $\Sigma$ by $\gamma_1\equiv-\Sigma$ \cite{NegeleOrland}. This allows for the calculation of the full Green's function by Dyson's equation
	\begin{align}\label{eq:Dyson}
		G_\text{Full}=\left(G_0^{-1}-\Sigma\right)^{-1}
	\end{align}
	Thus, we use $\Sigma\equiv-\gamma_1$ in the following and also rename the two particle vertex function as $\Gamma\equiv\gamma_2$ for consistency with previous literature.
	
	For the pseudo-fermion FRG the relevant parameters consist of $k=\{\omega_k,i_k,\beta_k,\alpha_k\}$ where $\omega$ is the frequency, $i$ is the lattice site, $\beta$ is the spin index and $\alpha$ is the Keldysh index. We assume that all one-particle quantities such as the self-energy $\Sigma$ and the Green's function $G$ only act locally, are diagonal in spin space and fulfill energy conservation. Using this we get
	\begin{align}
		\Sigma(1'|1)=\Sigma^{\alpha'_{1}\alpha^\noprime_{1}}(\omega_1)\delta(\omega^\noprime_{1}-\omega'_{1})\delta_{i'_{1}i^\noprime_{1}}\delta_{\beta'_{1}\beta^\noprime_{1}}.
	\end{align}
	The condition of locality is a consequence of the decomposition into pseudo fermions. This will be discussed in detail in Sec.~\ref{sec:symmetry}. For the two-particle functions one similarly gets a bi-locality condition. This allows for the vertex to be decomposed as
	
	\begin{align}\nonumber
		\Gamma&(1',2'|1,2)=\\
		\big\{&\,\big[\Gamma_{\mu\nu\,i_1i_2}^{\alpha'_{1}\alpha'_{2}\alpha^\noprime_{1}\alpha^\noprime_{2}}(\omega'_{1}\omega'_{2}|\omega^\noprime_{1}\omega^\noprime_{2})\sigma^\mu_{\beta'_{1}\beta^\noprime_{1}}\sigma^\nu_{\beta'_{2}\beta^\noprime_{2}}\nonumber\\
		+&\;\;\Gamma_{d\,i_1i_2}^{\alpha'_{1}\alpha'_{2}\alpha^\noprime_{1}\alpha^\noprime_{2}}(\omega'_{1}\omega'_{2}|\omega^\noprime_{1}\omega^\noprime_{2})\delta_{\beta'_{1}\beta^\noprime_{1}}\delta_{\beta'_{2}\beta^\noprime_{2}} \big]\delta_{i'_{1}i^\noprime_{1}}\delta_{i'_{2}i^\noprime_{2}}\nonumber\\
		-&\,\big[\Gamma_{\mu\nu\,i_2i_1}^{\alpha'_{1}\alpha'_{2}\alpha^\noprime_{2}\alpha^\noprime_{1}}(\omega'_{1}\omega'_{2}|\omega^\noprime_{2}\omega^\noprime_{1})\sigma^\mu_{\beta'_{1}\beta^\noprime_{2}}\sigma^\nu_{\beta'_{2}\beta^\noprime_{1}}\label{eq:vertexDecomp}\\
		+&\;\;\Gamma_{d\,i_2i_1}^{\alpha_{1'}\alpha_{2'}\alpha_{2^\noprime}\alpha_{1^\noprime}}(\omega'_{1}\omega'_{2}|\omega^\noprime_{2}\omega^\noprime_{1})\delta_{\beta'_{1}\beta^\noprime_{2}}\delta_{\beta'_{2}\beta^\noprime_{1}} \big]\delta_{i'_{1}i^\noprime_{2}}\delta_{i'_{2}i^\noprime_{1}}\big\}\nonumber\\
		\cross&\delta(\omega_1+\omega_2-\omega_{1'}-\omega_{2'})\nonumber.
	\end{align}
	Here, we used the fermionic properties under particle permutation and performed a basis decomposition in terms of spin components $\Gamma_{\mu\nu}$ and a density component $\Gamma_d$ using the Pauli matrices ($\sigma^\mu$ with $\mu\in{x,y,z}$) and the unit matrix ($\sigma^{\mu=0}\equiv\mathds{1}$). For an isotropic (Heisenberg) spin interaction one can assume $\Gamma_{\mu\nu}=\Gamma_s\delta_{\mu\nu} ,\,\forall\,\mu,\nu$ which further simplifies the decomposition. The anisotropic Heisenberg model with $\Gamma_{\mu\nu}=\Gamma_\mu\delta_{\mu\nu}$ also includes the XXZ model or the Kitaev-Heisenberg model. The full flow equations for the anisotropic vertices can be found in the supplemental material \cite{NoteSM}.
	
	The initial conditions in the Keldysh basis are given by
	\begin{align}
		\Gamma_{\mu\nu\,i_1i_2}^{\alpha'_{1}\alpha'_{2}\alpha^\noprime_{1}\alpha^\noprime_{2},\Lambda\to\infty}=-\I\frac{J^{\mu\nu}_{ij}}{4}\bar{v}^{\alpha'_{1}\alpha'_{2}\alpha^\noprime_1\alpha^\noprime_2}
	\end{align}
	where only the spin vertices are initialized and the Keldysh structure is given by~\cite{JakobsSIAM}
	\begin{align}
		\bar{v}^{1'2'12}=\begin{cases}
			\frac{1}{2} \!&\!\!\! (1'=1\land 2'\neq 2) \lor (1'\neq1\land 2'=2) \\
			0 & \!\!\! \text{otherwise.}
		\end{cases}
	\end{align}
	\subsection{Large-S Limit}\label{sec:larges}
	There are different limits in which the FRG equations simplify into a system which is easier to solve \cite{LargeS,DissBuessen}. One of these is the large $S$ limit, where the spin length of the localized moments goes to infinity. This renders the system classical in the sense, that the spins act like classical vectors, as the commutators scale with $1/S$ and thus vanish for $S\to\infty$~\cite{millard1971infinite,lieb1973classical}. We build the spin $S$ operators as a sum of $2S$ $S=1/2$ spins $S^\text{Full}_i=\sum_{k}S_{ik}$, where $k$ is an additional flavor, enumerating the spin $1/2$ operators. It was shown, that -- even without a repulsion term to favor longer spin lengths -- this choice creates the maximal spin $S$ ground state, albeit all other possibilities of the spin addition are possible \cite{LargeS,DissBuessen}.
    In this limit with the decomposition above, only the second line in Eq.~\eqref{eq:flowEq} is non-zero, as a result of the non-locality in the summation ($\sum_k \gamma_{ik}\gamma_{kj}$ vs $\gamma_{ij}\gamma_{ij}$) \cite{LargeS}.
    
    This non-local term is -- diagrammatically written -- a typical RPA-like ladder and can be solved directly by utilizing the geometric series. This is possible since the self-energy flow vanishes in this limit.
	
	Formally, the full two-particle vertex in this limit can be written as
	\begin{align}\label{eq:RPADecomp}
		\Gamma_{\text{RPA}}=\Gamma_\text{bare}+\Gamma_t.
	\end{align}
	Using the now exact Bethe-Salpeter equation
	\begin{align}
		\Gamma_t=\frac{1}{\pi}\sumint \mathrm{d}\Omega\; G\Gamma_\text{bare}\Gamma_\text{RPA}G,
	\end{align}
	where we denote all internal summations and integrations by $\sumint \mathrm{d}\Omega$ for brevity. This leads to the infinite ladder, when inserting into Eq.~\eqref{eq:RPADecomp}. The solution is given by
	\begin{align}
		\Gamma_{i_1i_2,\text{RPA}}^{1'2'1^\noprime2^\noprime}(t)=\sum_j\sum_{34'}\Gamma_{i_1j,\text{bare}}^{1'4'1^\noprime3^\noprime}R^{3^\noprime2'|4'2^\noprime}_{ji_2}(t),
	\end{align}
	where $R$ is calculated via
	\begin{align}\label{eq:RPASum}
		R\equiv\!\bigg(\!\mathds{1}^{32'4'2}_{ji_2}-\frac{1}{\pi}\sum_{3'4}\Gamma_{ji_2,\text{bare}}^{3'2'42}\!\int\!\mathrm{d}\nu \,G^\Lambda_{33'}(t+\nu)G^\Lambda_{44'}(\nu)\!\bigg)^{-1},
	\end{align}
	which is the result of the geometric series. The inversion can be performed using standard matrix inversion for $m_{ij}$ by rearranging the indices so that all summation indices ($j,3,4'$) are found in $i$ and the outside indices are in $j$. This assures, that $m^k$ is well-defined by standard matrix multiplication and thus also $R=m^{-1}$. Thus, we can calculate the irreducible vertex $\Gamma$ for the physical system with $\Lambda=0$, because the formulation of the RPA does not need an energy cutoff. However, this results in integrals with peaked integrands, which is why we introduce a finite $\Lambda$ to evaluate the RPA. Further, introducing the cutoff into the RPA allows us to compare the FRG results with the exact RPA results, to check whether the numerical implementation performs at the desired accuracy. This will be discussed in Sec.~\ref{sec:DimerRPA}. 
	
	\subsection{Symmetry Analysis}\label{sec:symmetry}
	An efficient numerical implementation of Keldysh PFFRG will necessitate the implementation of all possible symmetries of the system which are either due to the chosen model or general symmetries of the formalism. 
	We here bring together the general symmetries in the PFFRG~\cite{PFFRGReview} with the specific formulations for the Keldysh formalism, as laid out by Jakobs~\cite{JakobsSymmetries}.
	\subsubsection{General Symmetries}
	We start by the general symmetries of the formalism. For this we define a general $n$-particle Green's function as
	\begin{align}\label{eq:defGF}
		\I^nG^{\bm{\eta}|\bm{\eta'}}_{\bm{q}|\bm{q'}}(\bm{t}|\bm{t'})=\langle\mathcal{C}f_{q_1}^{\eta_1\ann}(t_1)\dots f_{q_n}^{\eta_n\ann}(t_n)f_{q'_{n}}^{\eta'_{n}\dagger}(t'_{n})\dots f_{q'_{1}}^{\eta'_{1}\dagger}(t'_{1})\rangle,
	\end{align}
	where the bold variables denote all of the $n$ corresponding variables and $q$ represents all necessary quantum numbers except the Keldysh indices and the time/frequency (i.e., the lattice site or the spin index). The Keldysh indices are denoted as $\eta=\pm$ in the contour and as $\alpha=1,2$ in the Keldysh basis, respectively. For better readability we separate the indices of incoming and outgoing particles explicitly (e.g., $\eta|\eta'$). For the Fourier transformation into frequency space we use the convention from Eqs.~\eqref{eq:FourierToOmega} and \eqref{eq:FourierToTime}.
	
	The first symmetry we consider already featured in Sec.~\ref{sec:Keldysh}, where we introduced the Keldysh rotation. If one time argument is strictly greater than all others, the contour index for this particle is ill-defined because the contour was artificially extended to $t\to\infty$ from the largest time. This is exploited by the Keldysh rotation. Applying the Keldysh rotation only to the contour index with the largest time yields
	\begin{align}
		G^{1,\eta_2\dots\eta_n|\bm{\eta'}}_{\bm{q}|\bm{q'}}(\bm{t}|\bm{t'})=0\quad\text{if } t_1>t_2\dots t_{n'}.
	\end{align}
	This means that in frequency space the Green's function with all Keldysh indices $\alpha_i=1$ fulfills
	\begin{align}
		G^{1\dots1|1\dots1}_{\bm{q}|\bm{q'}}(\bm{\omega}|\bm{\omega'})=0.
	\end{align}
	For the vertex functions there exists a similar relation~\cite{DissJakobs} 
	\begin{align}
		\gamma^{2\dots2|2\dots2}_{\bm{q}|\bm{q'}}(\bm{\omega}|\bm{\omega'})=0.
	\end{align} 
	This can be easily understood by Dyson's equation Eq.~\eqref{eq:Dyson} and the relation between the vertex and Green's functions.
	
	Next, we turn to the transformation of Green's functions under complex conjugation, which can directly be inferred from Eq.~\eqref{eq:defGF} as
	\begin{align}
		G^{\bm{\eta}|\bm{\eta'}}_{\bm{q}|\bm{q'}}(\bm{t}|\bm{t'})^*=(-1)^nG^{\bm{\bar\eta'}|\bm{\bar\eta}}_{\bm{q'}|\bm{q}}(\bm{t'}|\bm{t}),
	\end{align}
	where the bar denotes a switch in the contour from one branch to the other ($\pm\to\mp$). The pre-factor is due to the imaginary units in the definition of the Green's function. We do not get additional minus signs from contour ordering the operators, since they always come in pairs.
	For the Keldysh basis in frequency space the same relation is given by
	\begin{align}
		G^{\bm{\alpha}|\bm{\alpha'}}_{\bm{q}|\bm{q'}}(\bm{\omega}|\bm{\omega'})^*=(-1)^{n+\sum_k(\alpha_k+\alpha_{k'})}G^{\bm{\alpha'}|\bm{\alpha}}_{\bm{q'}|\bm{q}}(\bm{\omega'}|\bm{\omega}),
	\end{align}
	where the exchange of the branch now only results in an additional pre-factor and leaves the indices unchanged.
	Since the vertex functions are closely related to the Green's functions due to the tree expansion and Dyson's equation, all of the remaining symmetries are also applicable to the vertex functions while sometimes the pre-factors are slightly different \cite{DissBuessen,DissJakobs}. We thus only use the Green's functions in the following.
	
	The third symmetry is invariance under global time translations, which originates in the time-independence of the equilibrium Hamiltonian~\eqref{eq:HamPF}. This allows us to always perform a global time translation of our operators resulting in an energy conservation relation in frequency space
	\begin{align}
		G^{\bm{\eta}|\bm{\eta'}}_{\bm{q}|\bm{q'}}(\bm{\omega}|\bm{\omega'})=G^{\bm{\eta}|\bm{\eta'}}_{\bm{q}|\bm{q'}}(\bm{\omega}|\bm{\omega'})\delta\left(\sum_{k=1}^n(\omega'_k-\omega^\noprime_k)\right).
	\end{align}

	\begin{table*}
		\centering
		\begin{tabular*}{0.85\linewidth}{@{\extracolsep{\fill}}lc}
			\hline
			\hline
			$\Gamma^{2222}(1'2'|12)=0$&(CS)\\
			$\Gamma^{\alpha'_1\alpha'_2|\alpha^\noprime_1\alpha^\noprime_2}(1'2'|12)=(-1)^{\sum_k(\alpha^\noprime_k+\alpha_k')}\Gamma^{\alpha^\noprime_1\alpha^\noprime_2|\alpha'_1\alpha'_2}(12|1'2')^*$&(CC)\\
			$\Gamma^{\alpha'_1\alpha'_2|\alpha^\noprime_1\alpha^\noprime_2}(1'2'|12)=\Gamma^{\alpha'_1\alpha'_2|\alpha^\noprime_1\alpha^\noprime_2}(1'2'|12)\delta(\omega'_1+\omega'_2-\omega^\noprime_1-\omega^\noprime_2)$&(TT)\\
			$\Gamma^{\alpha'_1\alpha'_2|\alpha^\noprime_1\alpha^\noprime_2}(1'2'|12)=\Gamma^{\alpha'_1\alpha'_2|\alpha^\noprime_1\alpha^\noprime_2}(1'2'|12)\delta_{i^\noprime_1i'_1}\delta_{i^\noprime_2i'_2}-\Gamma^{\alpha'_1\alpha'_2|\alpha^\noprime_2\alpha^\noprime_1}(1'2'|21)\delta_{i^\noprime_2i'_1}\delta_{i^\noprime_1i'_2}$&(U(1))\\
			$\Gamma^{\alpha'_1\alpha'_2|\alpha^\noprime_1\alpha^\noprime_2}(1'2'|12)\delta_{i^\noprime_1i'_1}\delta_{i^\noprime_2i'_2}=-\Gamma^{\alpha'_1\alpha'_2|\alpha^\noprime_2\alpha^\noprime_1}(1'2'|21)\delta_{i^\noprime_2i'_1}\delta_{i^\noprime_1i'_2}=-\Gamma^{\alpha'_2\alpha'_1|\alpha^\noprime_1\alpha^\noprime_2}(2'1'|12)\delta_{i^\noprime_1i'_2}\delta_{i^\noprime_2i'_1}$&(X)\\
			$\Gamma^{\alpha'_1\alpha'_2|\alpha^\noprime_1\alpha^\noprime_2}(1'2'|12)\delta_{i^\noprime_1i'_1}\delta_{i^\noprime_2i'_2}=-\beta'_1\beta^\noprime_1\,\Gamma^{\alpha^\noprime_1\alpha'_2|\alpha'_1\alpha^\noprime_2}(-12'|\!-\!1'2)\delta_{i^\noprime_1i'_1}\delta_{i^\noprime_2i'_2}$&(PH1)\\
			$\Gamma^{\alpha'_1\alpha'_2|\alpha^\noprime_1\alpha^\noprime_2}(1'2'|12)\delta_{i^\noprime_1i'_1}\delta_{i^\noprime_2i'_2}=-\beta'_2\beta^\noprime_2\,\Gamma^{\alpha'_1\alpha^\noprime_2|\alpha^\noprime_1\alpha'_2}(1'\!\!-\!2|1\!-\!2')\delta_{i^\noprime_1i'_1}\delta_{i^\noprime_2i'_2}$&(PH2)\\
			\hline
			\hline
		\end{tabular*}  
		\caption{All symmetries of the two particle vertex function used in this work. We explicitly write out the Keldysh indices for all symmetries even if they remain unchanged. (CS) is the causality relation, (CC) is the symmetry under complex conjugation in the Keldysh basis, (TT) is the symmetry under global time translations, (U(1)) is the U(1) gauge freedom, (X) is the crossing symmetry and (PH1) and (PH2) denote the particle-hole conjugation in the first or second particle. The pre-factor for the (CC) symmetry misses a $-1$ compared with Ref. \cite{JakobsSymmetries} since the vertex definition contains an additional imaginary unit. }  
		\label{tab:symmetries}
	\end{table*}
	
	Furthermore, we have the crossing symmetry. Due to the fermionic nature of the operators we get 
	\begin{align}
		G^{\bm{\alpha}|P(\bm{\alpha'})}_{\bm{q}|P(\bm{q'})}(\bm{\omega}|P(\bm{\omega'}))&=G^{P(\bm{\alpha})|\bm{\alpha'}}_{P(\bm{q})|\bm{q'}}(P(\bm{\omega})|\bm{\omega'})\nonumber\\&=(-1)^P G^{\bm{\alpha}|\bm{\alpha'}}_{\bm{q}|\bm{q'}}(\bm{\omega}|\bm{\omega'}),
	\end{align}
	where $P$ denotes an arbitrary permutation and $(-1)^P$ is the resulting sign due to the total number of fermionic permutations.
	
	We can also make use of all symmetries, that map equivalent lattice points onto another. We denote those symmetry operations by $L(\bm{i})$. This leaves all other quantum numbers unchanged. We get
	\begin{align}
		G_{i|i'}(1|1')=G_{L(i)|L(i')}(1|1'),
	\end{align}
	and similarly for the two particle Green's functions and the vertex functions.
	
	Note that we will not use the symmetry under time reversal. This is due to the contour ordering of the operators, which would also be reversed under this operation, leading to anti-contour ordered Green's functions. In equilibrium one can use this to extract generalized fluctuation-dissipation theorems \cite{DissJakobs}, which we not use due to their complicated Keldysh structure, thus rendering a numerical implementation unfeasible~\footnote{If we stayed in the $\pm$ basis, the generalized FDT's would be easy to implement but the causality relations then are difficult. We decided to use the Keldysh basis, since the FDT's are only fulfilled in equilibrium and thus not as general as the causality relations.}. However, they can be applied as a check of the symmetry violation of the implementation.
	
	\subsubsection{PFFRG Specific Symmetries}
	The two specific symmetries result from the invariance under local SU(2) rotations of the pseudo-fermion decomposition~\eqref{eq:decomposition} and thus the Hamiltonian~\eqref{eq:HamPF}. 
	Since, a general SU(2) transformation would give a linear combination of creation and annihilation operators, the evaluation of the resulting expectation values would give a mixture of different operators. Due to this, we only choose the subgroups that have a one-to-one mapping in the operators. Those are the U(1) and the $\mathds{Z}_2$.
	
	The transformation of the pseudo-fermion operators under local U(1) operations can be written as
	\begin{align}
		g_{\varphi_i}\begin{pmatrix}
			f^{\eta\dagger}_{i\beta'}\\
			f^{\eta\ann}_{i\beta}
		\end{pmatrix}
		g^{-1}_{\varphi_i}=
		\begin{pmatrix}
			\mathrm{e}^{i\varphi_i}f^{\eta\dagger}_{i\beta'}\\
			\mathrm{e}^{-i\varphi_i}f^{\eta\ann}_{i\beta}
		\end{pmatrix},
	\end{align}
	which is just the multiplication by a phase $\varphi_i$ at lattice site $i$. For the one-particle Green's function, invariance under this transformation leads to the condition
	\begin{align}
		G_{i|i'}(1|1')\stackrel{!}{=}\mathrm{e}^{\I(\phi_{i'}-\phi_{i^\noprime})}G_{i|i'}(1|1').
	\end{align}
	Since, the phase is independent on all lattice sites, this fixes the one particle functions to be local and, likewise, all two particle functions to be bi-local.
	The $\mathds{Z}_2$ operations are the particle-hole conjugations
	\begin{align}
		g^{\vphantom{-1}}_\text{PH}\begin{pmatrix}
			f^{\eta\dagger}_{i\beta}\\
			f^{\eta\ann}_{i\beta}
		\end{pmatrix}
		g_\text{PH}^{-1}=
		\begin{pmatrix}
			\beta f^{\eta\ann}_{i\bar\beta}\\
			\beta f^{\eta\dagger}_{i\bar\beta}
		\end{pmatrix},
	\end{align}
	where $\bar\beta$ denotes a flipped spin index.
	Note, that this symmetry always has to be applied to both branches, thus allowing for the symmetry to be used for arbitrary Keldysh indices.
	
	For the one-particle Green's function this results in
	\begin{align}
		G^{\alpha|\alpha'}_{\beta|\beta'}(\omega|\omega')\delta_{ii'}=-\beta\beta'G^{\alpha'|\alpha}_{\bar{\beta}'|\bar{\beta}}(-\omega'|-\omega)\delta_{ii'}.
	\end{align}
	and can be cross checked easily for the free particle propagators.
	For the two-particle Green's function we get
	\begin{align}
		G^{\alpha^\noprime_1\alpha^\noprime_2|\alpha'_1\alpha'_2}_{\beta^\noprime_1\beta^\noprime_2\,|\beta'_1\beta'_2}&(\omega^\noprime_1,\omega^\noprime_2|\omega'_1,\omega'_2)\delta_{i^\noprime_1i'_1}\delta_{i^\noprime_2i'_2}\\&\!\!\!=-\beta^\noprime_1\beta'_1G^{\alpha'_1\alpha^\noprime_2|\alpha^\noprime_1\alpha'_2}_{\bar{\beta}'_1\beta^\noprime_2\,|\bar{\beta}^\noprime_1\beta'_2}(-\omega'_1,\omega^\noprime_2|\!-\!\omega^\noprime_1,\omega'_2)\delta_{i^\noprime_1i'_1}\delta_{i^\noprime_2i'_2}\nonumber\\
		&\!\!\!=-\beta^\noprime_2\beta'_2G^{\alpha^\noprime_1\alpha'_2|\alpha'_1\alpha^\noprime_2}_{\beta^\noprime_1\bar{\beta}'_2\,|\beta'_1\bar{\beta}^\noprime_2}(\omega^\noprime_1,-\omega'_2|\omega'_1,-\omega\noprime_2)\delta_{i^\noprime_1i'_1}\delta_{i^\noprime_2i'_2},\nonumber
	\end{align}
	where both corresponding particle-hole pairs can be mapped independently.
	
	A list of all vertex symmetries used in this paper is presented in Table \ref{tab:symmetries}. The combinations of symmetries used in the numerical implementation are given in the supplemental material \cite{NoteSM} for completeness.
	
	\subsection{Observables}\label{sec:observables}
	
	The PFFRG equations inherently respect all symmetries present in the initial conditions, i.e., the model as defined by the Hamiltonian. Therefore, no order parameters, such as a magnetization in the case of magnetic order, can become finite during the flow. To enter the symmetry broken phase during the RG procedure, one would have to include a small initial ordering tendency, which could be realized, e.g., through an infinitesimal magnetic field to probe for magnetic orders. If this initial order is accepted by the system, the corresponding magnetization then will become finite during the flow~\cite{PFFRGMagnetic,PMFRG_Magnetic}.
	This procedure, however, requires an \emph{a priori} guess for the ordering tendency, which introduces a bias in the method. Therefore we refrain from this technique and defer it to future research.
	
	Instead, we use the magnetic susceptibility and the corresponding spin structure factor as a measure for ordering tendencies. If the system wants to order during the flow from the initially paramagnetic phase, this is signaled by a divergence of the magnetic susceptibility and breakdown of the RG flow. A lack of such a divergence signals a magnetically disordered ground state and therefore a possible spin liquid candidate phase.
	
	To find an expression for the susceptibility in terms of pseudo fermions, we start by assuming that the action $\mathcal{S}$ in principle has two source terms for the spin along the $z$ direction  
	\begin{align}
		\mathcal{S}\propto h_+S^{z,+}-h_-S^{z,-},
	\end{align}
	since we can either put an operator on the $+$ or on the $-$ branch. The magnetic susceptibility in the contour basis then can be calculated using $\chi^{\pm\pm}=\partial \log(Z)/\partial h_\pm\partial h_\pm|_{h_\pm\to0}$ where $Z$ is the partition function. To get a physical result however, one has to calculate the retarded susceptibility, which can be done by either rotating the result of the susceptibility calculations afterwards, or by rotating directly in the action. We choose the latter variant. 
	
	After the decomposition using Eq.~\eqref{eq:decomposition} and rotation with $f^1~=~1/\sqrt{2}(f^+-f^-)$ and $f^2=1/\sqrt{2}(f^++f^-)$ (cf. Eq.~\eqref{eq:Keldysh rotation}) we get
	\begin{align}\nonumber
		\mathcal{S}_\text{mag}\propto\sum_{\alpha\alpha'\beta\beta'}&h_1(f^{1\dagger}_{\alpha'}\sigma^z_{\alpha'\alpha}f^{1\ann}_{\alpha\noprime}+f^{2\dagger}_{\alpha'}\sigma^z_{\alpha'\alpha}f^{2\ann}_{\alpha\noprime})\\
		+&h_2(f^{1\dagger}_{\beta'}\sigma^z_{\beta'\beta}f^2_{\beta}+f^{2\dagger}_{\beta'}\sigma^z_{\beta'\beta}f^1_{\beta})\\
		\equiv\sum_{\alpha\alpha'\beta\beta'}&V_q\rho_{cl}+V_{cl}\rho_q\nonumber.
	\end{align}
	Here, the last line represents the standard naming for an arbitrary field with sources $V$~\cite{Kamenev} with $V_q\equiv h_1$ and $V_{cl}\equiv h_2$. The real space spin susceptibility can then be calculated by
	
	\begin{align}\label{eq:susceptibility}
		\chi^\text{Ret}_{ij}&=\expval{S^{z}_{i}S^{z}_{j}}_\text{ret}=\frac{\delta \rho_{cl}}{\delta V_{cl}}=\frac{1}{2\I}\frac{\delta^2\log(Z)}{\delta V_{cl}\delta V_q}\\
		&=\frac{\I}{2}\!\!\!\!\sum_{\alpha\alpha'\beta\beta'}\!\!\!\sigma^z_{\alpha'\alpha}\sigma^z_{\beta'\beta}\big(G^{12|11}_\text{Full}\!+G^{11|12}_\text{Full}\!+G^{22|21}_\text{Full}\!+G^{21|22}_\text{Full}\big)\nonumber,
	\end{align}
	where we used the definition of the full Green's function $\I^2G^{12|1'2'}_\text{Full}=\expval{f^{1\ann} f^{2\ann} f^{2'\cre} f^{1'\cre}}$. Since the FRG only calculates the irreducible vertex functions which in turn allow to calculate connected quantities, we have expanded the full Green's function in terms of connected ones~\cite{NegeleOrland}
	\begin{align}\label{eq:GFdecomposition}
		G_\text{Full}^{12|1'2'}=G_c^{12|1'2'}+G_c^{1|1'}G_c^{2|2'}- G_c^{1|2'}G_c^{2|1'}.
	\end{align}
	These connected Green's functions need further decomposition into the irreducible two-particle vertex function and one-particle Green's functions using the so called tree expansion. The full decomposition is shown in the supplemental material\cite{NoteSM}.
	
	The dynamical spin structure factor is calculated from the imaginary part of the spin susceptibility via the fluctuation-dissipation theorem as
	\begin{align}\label{eq:structreFactor}
		S^\text{Ret}_{ij}(\omega)=\frac{1}{\pi}(1-\mathrm{e}^{-\beta\omega})^{-1}\text{Im}(\chi_{ij}^\text{Ret}(\omega)).
	\end{align}
	With this we are able to compare our results to inelastic neutron scattering experiments because they are able to directly measure the dynamic spin structure factor.
	
	\subsection{Numerical Implementation}\label{sec:numerics}
	For the implementation of the flow equation we take a route which is similar to state-of-the-art Matsubara implementations \cite{Multiloop,DissReuther,DissBuessen,PFFRGMagnetic,Benchmark}. That is, calculating the right hand side of the differential equation exploiting as many symmetries as possible while retaining structures with good ability to be vectorized by the compiler. This means, that we reduce the real space lattice to a minimum set of necessary symmetry in-equivalent points~\cite{DissReuther} and use the different symmetries from Sec.~\ref{sec:symmetry} to remove all negative frequencies by mapping to corresponding Keldysh indices and vertices. The typical approach in PFFRG is to assume an infinite lattice and truncate at a certain interaction distance by setting $\Gamma_{i,j}=0$ for $\norm{i-j}>d_\text{max}$~\footnote{Note, that this choice of lattice cutoff is not possible for direct RPA calculations, where the inversion in \cref{eq:RPASum} is only possible for a finite lattice.}.
	
	For the frequency discretization of the two particle vertex we use a linear-logarithmic grid with either 42 or 50 positive frequencies and one additional point at $\omega=0$. The number of grid points is chosen such that a further increment has no significant impact on the result. The discretization of the self-energy is chosen to be 10 times finer, since the numerical performance loss is insignificant.
	Furthermore, we use the asymptotic frequency parametrization~\cite{AsymptoticParametrization} to correctly capture the asymptotic behavior for the different diagrams. All points between grid points are multi-linearly interpolated. To avoid interpolation artifacts, all asymptotic classes have the same frequency discretization.
	Additionally, we perform a frequency re-meshing of all quantities up to some given cutoff $\Lambda_\text{min}$ using the same interpolation to accurately capture both the large frequency structures at high cutoffs as well as the small frequency features at the end of the flow. The stop of the re-meshing procedure is supported by the re-normalizing behavior of the self-energy, which ensures, that propagator features do not become infinitely sharp. Due to the logarithmic nature of the mesh, this re-meshing sometimes leads to high frequency artifacts, which are also apparent in the RPA analysis shown in the supplemental material \cite{NoteSM}, but are not significant for the computation of the observables.
	
	For solving the differential equations we use an error-controlled adaptive third order Adams-Bashforth-Moulton stepper~\cite{ODE} which we chose because of the superior performance for FRG in comparison to other steppers~\cite{BetterIntegrators}. For the frequency integration we opted for a standard adaptive Simpson rule~\cite{SimpsonsRule} with Richardson extrapolation~\cite{RichardsonExtrapolation} where we explicitly provide important integration points.
	
	The temperature is implemented into the Fermi distribution which only enters the free Keldysh propagator. This has the consequence that switching from $T=0$ to $T>0$ is far simpler compared with the Matsubara formalism, where the frequency integration at $T=0$ transforms into a discrete sum for $T>0$. The temperature behavior will be discussed in the next section.

    \subsubsection{Keldysh vs. Matsubara}
    For a better understanding of the differences between a Keldysh and a Matsubara implementation of the PFFRG, we highlight the key differences and also discuss how they would be implemented in an existing Matsubara PFFRG code. 
    First, due to the distinct behavior of time-reversal symmetry on the Keldysh contour, the combination of complex conjugation and time reversal ((CC) $\circ$ (T)) can no longer be used to enforce the Green's and vertex functions to be purely real functions as in the Matsubara formalism. 
    As a result, all propagators and vertices must be treated as complex-valued quantities ($\in \mathds{C}$). This necessitates a fundamental change in data structures: for example, in \texttt{C++}, one must replace \texttt{std::double} with \texttt{std::complex<double>} or an equivalent structure with proper complex arithmetic.
    Additionally, the Keldysh indices as well as their internal summation has to be implemented for the propagators and vertex functions. Due to causality, the self-energy and the free propagators needs to have three components and the vertex functions increase to 15 components per frequency discretization point. Due to the large number of vertex functions it is advised to access the stored data in bulk and perform the summation afterwards, so the compiler is able to parallelize the required steps. 
   This structural shift also affects numerical integration. Standard library (STL) adaptive integrators typically lack support for complex numbers or multi-function arrays. The latter is especially restrictive, as adaptive behavior then requires all functions to share the same integration grid. To address this, we implemented a simple yet reliable third-order adaptive integrator tailored for parallel integration of large arrays of functions. While not optimal in general performance, it performs well in this specific context.

    Concerning the cutoff, we advise to use the cutoff described above, as it does not violate the causality relations, which otherwise would add an additional source of error into the implementation. A more detailed comparison can be found in the Supplemental Material \cite{NoteSM}.

	\begin{figure}[t]
		\centering
		\includegraphics[width=.99\linewidth]{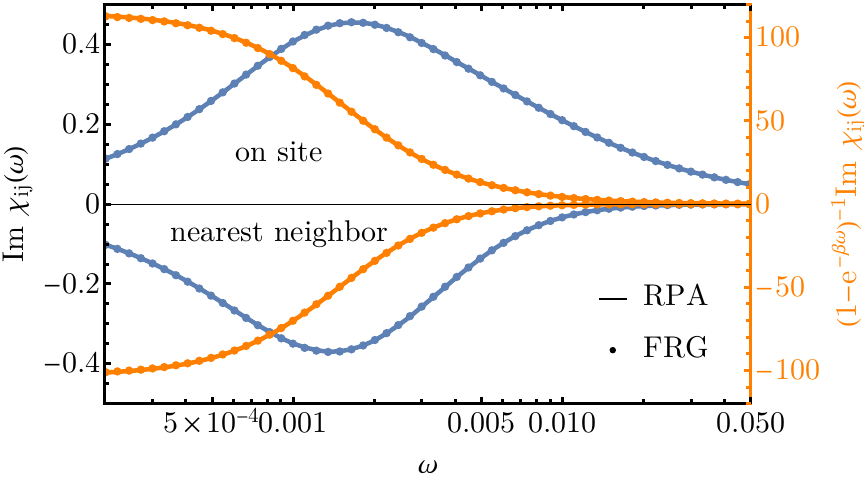}
		\caption{Dynamic susceptibility and resulting spin structure factor [from \cref{eq:structreFactor}] for the antiferromagnetic dimer in the $S\to\infty$ limit at $T=0.2J$. The cutoff is chosen such that $\Lambda=0.002J\ll T$. The results of the FRG coincide with the single step RPA calculation. The spin structure factor shows the expected peak around $\omega=0$, falling off relatively quickly.
		}
		\label{fig:RPADimer}
	\end{figure}
	
	\begin{figure*}[t!]
		\includegraphics[width=.49\linewidth]{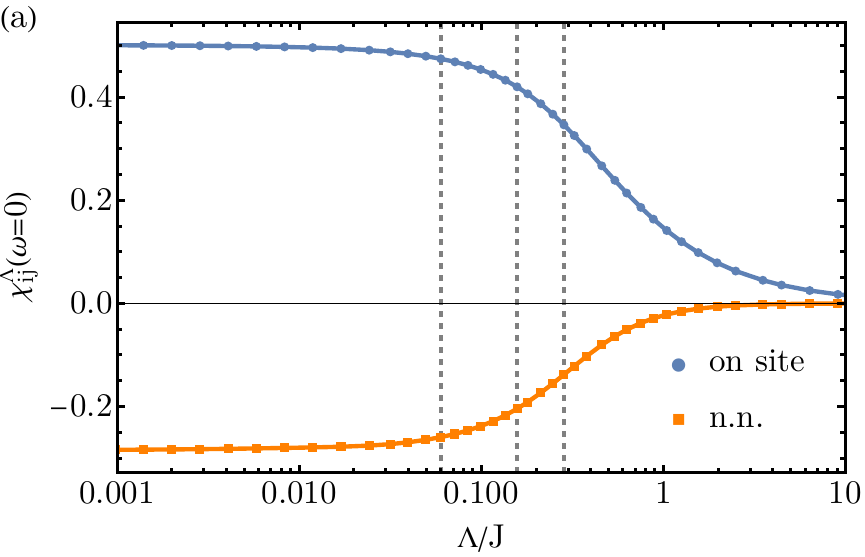}
		\includegraphics[width=.48\linewidth]{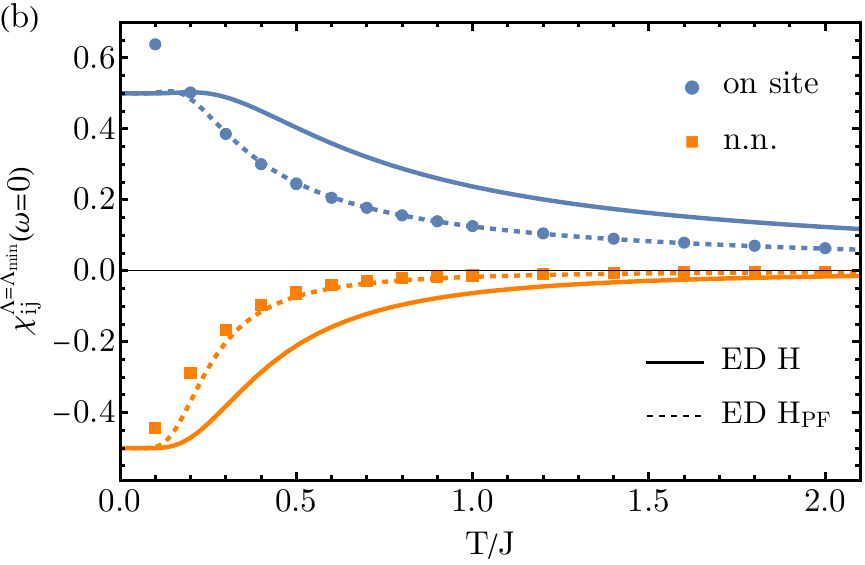}
		\label{fig:TempScalingDimer}
		\caption{(a) Exemplary FRG flow at $T=0.2 J$ for the antiferromagnetic Heisenberg dimer. The flow exhibits a saturation around $\Lambda=0.001J$ but is continued down to $\Lambda_\text{min}=10^{-6}J$ to exclude any influence of the cutoff scale. The dashed lines show the cutoffs, which are used for the comparison in \cref{fig:DimerSus}. (b) Temperature scaling of the susceptibilities at lowest cutoff $\Lambda_\text{min}$. The Keldysh PFFRG results match the exact diagonalization results (ED) of the pseudo-fermion dimer down to $T\approx0.2J$. For smaller $T$ the FRG overestimates the susceptibilities.
		}
		\label{fig:DimerMisc}
	\end{figure*}
	
	\section{Results}\label{sec:results}
	To test our technique we analyze a multitude of models for benchmarking purposes. We thus focus on the lower dimensional systems with spatial dimension $D\leq2$ as there are many numerical, experimental and exact results we can cross-check our results with. This will demonstrate the reliability of the Keldysh PFFRG implementation in comparison to other numerical techniques.
	
	\subsection{Dimer}\label{sec:dimer}
	
	Since the antiferromagnetic dimer has evolved to be the standard benchmark for PFFRG and the pseudo-Majorana FRG (PMFRG)~\cite{PseudoMajorana1,TemperatureFlow,PFFRGReview}, we use it to compare our technique to the currently available results and extend them to finite frequencies. For this we start with the easier limit, which is $S\to\infty$, and benchmark against the semi-analytical RPA results. After that, we focus on the full FRG results in the Keldysh formalism and the improvement on the RPA.

	\subsubsection{RPA Results}\label{sec:DimerRPA}
	The dimer is a finite size system both in FRG and RPA. Thus, the results of the FRG at a given cutoff should exactly coincide with the RPA results. The latter are calculated semi-analytically by performing frequency integrals and subsequently inverting the results according to~\cref{eq:RPASum}. Since, the RPA calculation scales only linearly with the frequency discretization, we can employ a much finer frequency resolution compared with the FRG approach.
	
	Due to the missing self-energy in the RPA formalism, the temperature plays a crucial role in smoothening the peaked structures. This is necessary, as the frequency resolution would have to be increased significantly to avoid a divergence in the flow for small $T$ with $\Lambda\ll T$ and still resolve all details correctly. In the following we choose $\beta=5/J$ ($T=0.2J$) because it is sufficient to show all relevant features when drawing a comparison. Smaller temperatures would necessitate a finer resolution in the FRG part of the calculation, which would be possible in this case but not for the remainder of the paper.
	
	The dynamic susceptibility and the resulting spin structure factor are shown in \cref{fig:RPADimer}. At finite temperatures the dynamic susceptibility is constrained to zero by symmetry at $\omega=0$, leading to peaks at finite $\omega$. To arrive at a less biased quantity we calculate the spin structure factor according to \cref{eq:structreFactor} which leads to the expected behavior around $\omega=0$, since the classical $S\to\infty$ limit has gapless excitations and thus a peaked response at $\omega=0$.
	
	Comparing the FRG results with the RPA data, we can see that the FRG results coincide with the much more accurate RPA scheme. The relative deviations are below $0.3\%$, which is negligible in comparison to other systematic deviations of the method. A comparison on vertex level further underlines a good performance of the FRG as shown in the supplemental material \cite{NoteSM}. Thus, we can verify that the FRG is able to resolve the main features during the flow and the frequency resolution of $N_\omega=50$ is sufficient to approximate the RPA results with $N_\omega=300$.
	
	\begin{figure*}[t]
		\includegraphics[width=.48\linewidth]{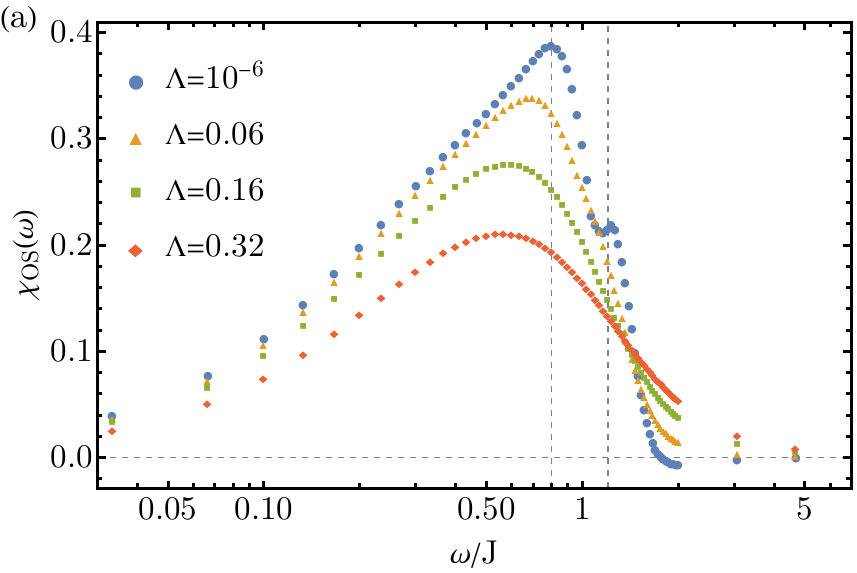}
		\label{fig:sfig1}
		\includegraphics[width=.48\linewidth]{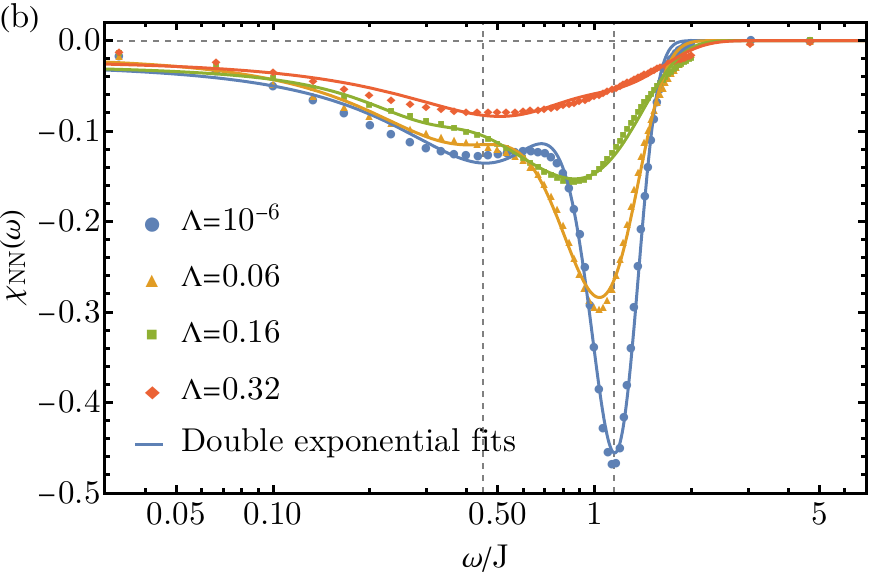}
		\label{fig:sfig2}
		\caption{Frequency resolved susceptibilities for the AFM dimer at $T=0.2J$ for different points in the flow. The chosen cutoffs are also marked in Fig.~\ref{fig:DimerMisc}(a).
			In panel (b) we also show double exponential fits according to Eq.~\eqref{eq:doubleExponential}, which describe the susceptibilities up to some minor differences. 
			The peak positions for the smallest cutoffs are marked by dashed lines for better visibility.
		}
		\label{fig:DimerSus}
	\end{figure*}
	
	\subsubsection{FRG Results}
	Following the RPA comparison we now turn to the quantum $S=1/2$ dimer, where we cannot neglect the remaining parts of the vertex flow and the self-energy also becomes finite. However, due to its simple structure, the antiferromagnetic dimer is exactly solvable, which allows for a good benchmark of our technique. Since, the FRG calculates susceptibilities, we compare against the exact results for both the physical system ($\chi_{\text{os/nn}}$ for on-site/nearest-neighbor susceptibilities, respectively) as well as the dimer in pseudo-fermion decomposition without projection to the physical subspace ($\chi_{\text{os/nn}}^\text{pf}$). This allows us to assess the effect of the pseudo-fermion constraint on our analysis. The exact susceptibilities for $\omega=0$ are given by
	
	\begin{gather}\label{eq:SusExact}
		\setlength{\displaywidth}{\columnwidth-100pt}
		\chi_{\text{os}}(\beta)=\frac{\beta+\mathrm{e}^{\beta}-1}{2\mathrm{e}^{\beta/4}Z_\text{ph}(\beta)},\quad \chi_{\text{nn}}(\beta)=\frac{\beta-\mathrm{e}^{\beta}+1}{2\mathrm{e}^{\beta/4}Z_\text{ph}(\beta)},\\ \label{eq:SusPF}
		\chi_{\text{os}}^\text{pf}(\beta)=\frac{\beta+\mathrm{e}^{\beta}-1+2\beta\mathrm{e}^{\beta/4}}{2\mathrm{e}^{\beta/4}Z_\text{pf}(\beta)},\quad\chi_{\text{nn}}^\text{pf}(\beta)=\frac{\beta-\mathrm{e}^{\beta}+1}{2\mathrm{e}^{\beta/4}Z_\text{pf}(\beta)},\nonumber\\
		\mathrm{e}^{\beta/4}Z_\text{ph}(\beta)=\mathrm{e}^{\beta}+3,\quad \mathrm{e}^{\beta/4}Z_\text{pf}(\beta)=\mathrm{e}^{\beta}+3+12\mathrm{e}^{\beta/4}.\nonumber
	\end{gather}
	These quantities can be calculated using the Lehmann representation \cite{PseudoMajorana1} which is laid out in the supplemental material \cite{NoteSM}.
	
	We perform the FRG calculations with $N_\omega=50$ for different temperatures. For $T>0.1J$ we get a monotonic flow down to the smallest cutoff $\Lambda$, which we set to $\Lambda_\text{min}=10^{-6}J$ to avoid any effects that a finite cutoff could have on the results. This can be done in the FRG since the self-energy regularizes the divergence and, thus, sharp peaks, which appeared in the RPA, are avoided. An exemplary flow of the on-site and nearest-neighbor susceptibility at $\omega=0$ is depicted in Fig.~\ref{fig:DimerMisc}(a). It illustrates that the results are already cutoff independent for $\Lambda<0.01J$ as the flow saturates. 
	
	The final values of the susceptibilities for each temperature are depicted in Fig.~\ref{fig:DimerMisc}(b). It shows that for $T>0.2J$ the results match with the exact pseudo-fermion solution but not the exact dimer. This discrepancy arises because we only assume the constraint Eq.~{\eqref{eq:PFConstraint}} to be fulfilled on average. Therefore, we allow the unphysical states to enter the results. As can be seen from Eqs.~\eqref{eq:SusExact} these mainly results in the change of the partition function by a constant term. Only in the on-site susceptibility there is a possibility for an unphysical state to play a role in the expectation value, but this is also only for $\omega=0$. Thus, we would assume that the pseudo-fermion constraint does not impact the results more than change their scaling. Unfortunately, this reasoning is only valid for a FRG without truncation, as the truncation in principle allows for the unphysical behavior to be enhanced. This issue was already discussed by Schneider {\it et.~al.}~\cite{BenediktPopov} who used the Popov-Fedotov technique~\cite{Popov,Popov2} to project out the unphysical states in the partition function by applying an imaginary chemical potential. However, this removes the symmetry of the vertex under complex conjugation and thus further increases the numerical effort. As the Keldysh PFFRG is already numerically expensive, we decided against employing this technique in this study. A more suitable approach would be to implement the Keldysh formalism into the pseudo Majorana FRG as it does not have the drawback of violating the particle number constraint. 
	
	Moreover, the FRG tends to exhibit ordering behavior, even in scenarios where it would be prohibited by the Mermin-Wagner theorem. This phenomenon is observed for the dimer at $T=0.1J$, where the flow shows a small plateau in the nearest neighbor susceptibility instead of remaining monotonous [see supplemental material \cite{NoteSM}]. This suggests that the system is attempting to transition away from the disordered paramagnetic starting state, which in turn renders the FRG flow unstable. Consequently, we refrain from continuing the analysis below $T=0.1J$ for the dimer.
	
	Additional deviations can be attributed to the truncation of the FRG equations. In our analysis, we truncated at the one-loop level, which effectively neglects three-particle contributions. The Katanin substitution \cite{Katanin} provides partial compensation for this omission, but higher-order terms still remain unaccounted for in the FRG flow. Unfortunately, this truncation can introduce uncontrollable effects, leading to a flow that is cutoff-dependent rather than cutoff-universal~\cite{DissJakobs}. As a result, using the FRG as a quantitative tool becomes problematic when higher-order terms cannot be excluded. Although extending the method to include higher-loop terms appears promising, the benefits may not justify the exponentially increased computational cost~\cite{Multiloop,thoennissmultiloop,Kuglermultiloop}. Nonetheless, the qualitative predictions of PFFRG have already proven reliable for one-loop calculations, with increasing accuracy in higher-dimensional systems \cite{PFFRGReview,Chillal-2020,Gonzalez-2024,Niggemann-2023,Noculak-2023,Gomez-2024,Gresista-2025,Ghosh-2019,Ghosh-2019a,Iqbal-2019,Iqbal-2018_diamond,Iqbal-2017,Iqbal-2016}. The zero-dimensional dimer, therefore, serves as a benchmark with the potentially severest qualitative issues.
	
	Unphysical excitations are also apparent in the frequency spectrum of both in-equivalent dimer sites, as illustrated in Fig.~\ref{fig:DimerSus} for $T=0.2J$. A double peak structure is observed in both the on-site and nearest neighbor susceptibilities. However, from the exact result
	\begin{align}
		\chi_\text{os/nn}^{(\text{pf})}(\omega)\propto\lim_{\eta\to0}\pm\left(\frac{1}{\omega+\mathrm{i}\eta-J}-\frac{1}{\omega+\mathrm{i}\eta+J}\right),
	\end{align}
	we would expect only a single peak for $\omega>0$ in the imaginary part of the susceptibility exactly at $\omega=J$, as this is the energy required to induce a spin flip from the ground state. Thus, we identify the largest peak in the nearest neighbor susceptibility as the physical excitation.
	
	We perform a fit of the nearest-neighbor susceptibility according to 
	\begin{align}
		\chi_{\text{NN,Fit}}(\omega)=A_1\mathrm{e}^{-\frac{1}{2}\frac{(\omega-\omega_1)^2}{\sigma_1^2}}+A_2\mathrm{e}^{-\frac{1}{2}\frac{(\omega-\omega_2)^2}{\sigma_2^2}},\label{eq:doubleExponential}
	\end{align}
	
	also shown in \cref{fig:DimerSus}(b). 
	
	We find, that there is a deviation of the main peak from the expected value of $\omega = J$. We find, that the peak is at $\omega<J$ for both higher temperatures and cutoffs, shifting toward $J$ upon lowering any two of these, and eventually reaching $\omega>J$ for low temperatures and cutoffs. This deviation as well as the additional peak we attribute to the inherent approximations of PFFRG.
	As evident from the varying cutoff values $\Lambda$, the broadening is also significantly amplified by the cutoff scale. Therefore, only in the case of $\Lambda\ll T$ can we reasonably assume that the results do not exhibit any additional broadening beyond the truncation effects and temperature broadening. For the on-site susceptibility, a double exponential fit failed to accurately describe the data points.
	
	Based on this analysis of the dimer and the pseudo-fermion constraint, we limit our investigation to $T=0.1J$ as the lowest temperature value, aligning with typical PFFRG boundaries~\cite{BenediktPopov}, provided, the flow exhibits a monotonous behavior. However, selecting the smallest possible temperature is encouraged, as temperature broadening tends to obscure distinct features. Additionally, we acknowledge that certain modes may not exhibit the correct physical behavior, though we expect these modes to be small compared with the physically relevant ones, as seen for the nearest neighbor susceptibility.
	
	\subsection{1D Spin Chain}\label{sec:spinchain}
	
	\begin{figure}[t]
		\centering
		\includegraphics[width=.99\linewidth]{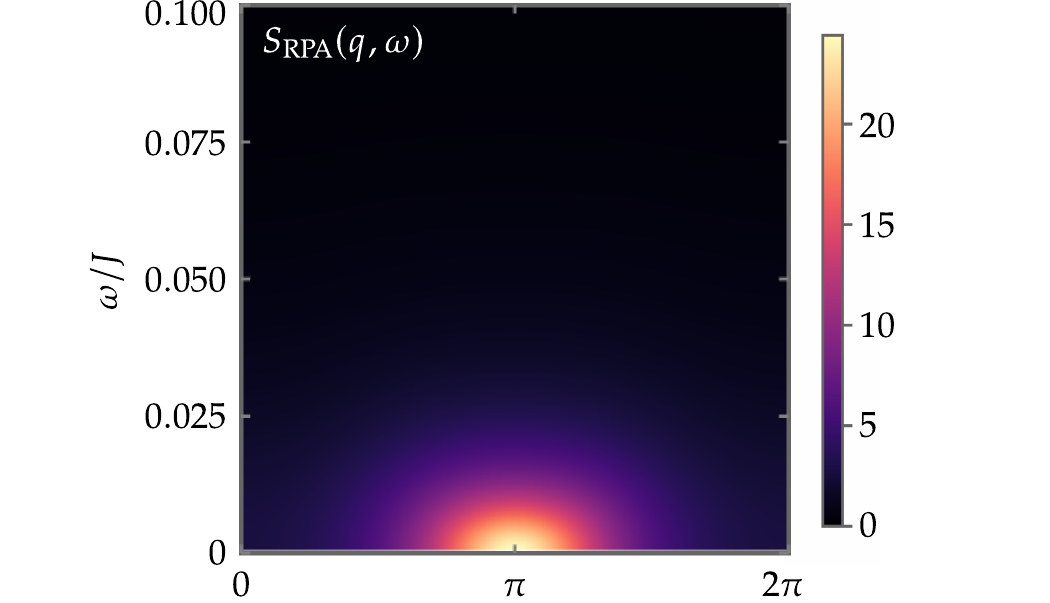}
		\caption{Spin structure factor for the spin chain with $N=10$ in the $S\to\infty$ limit. The excitation is localized around $q=\pi$ and falls off quickly for $\omega>0$. We do not get any branches, as we would get in spin wave theory since the initial state in the FRG is fully disordered and thus paramagnetic .
		}
		\label{fig:RPASpinChain}
	\end{figure}
	
	\begin{figure*}[t!]
		\centering
		\includegraphics[width=.99\linewidth]{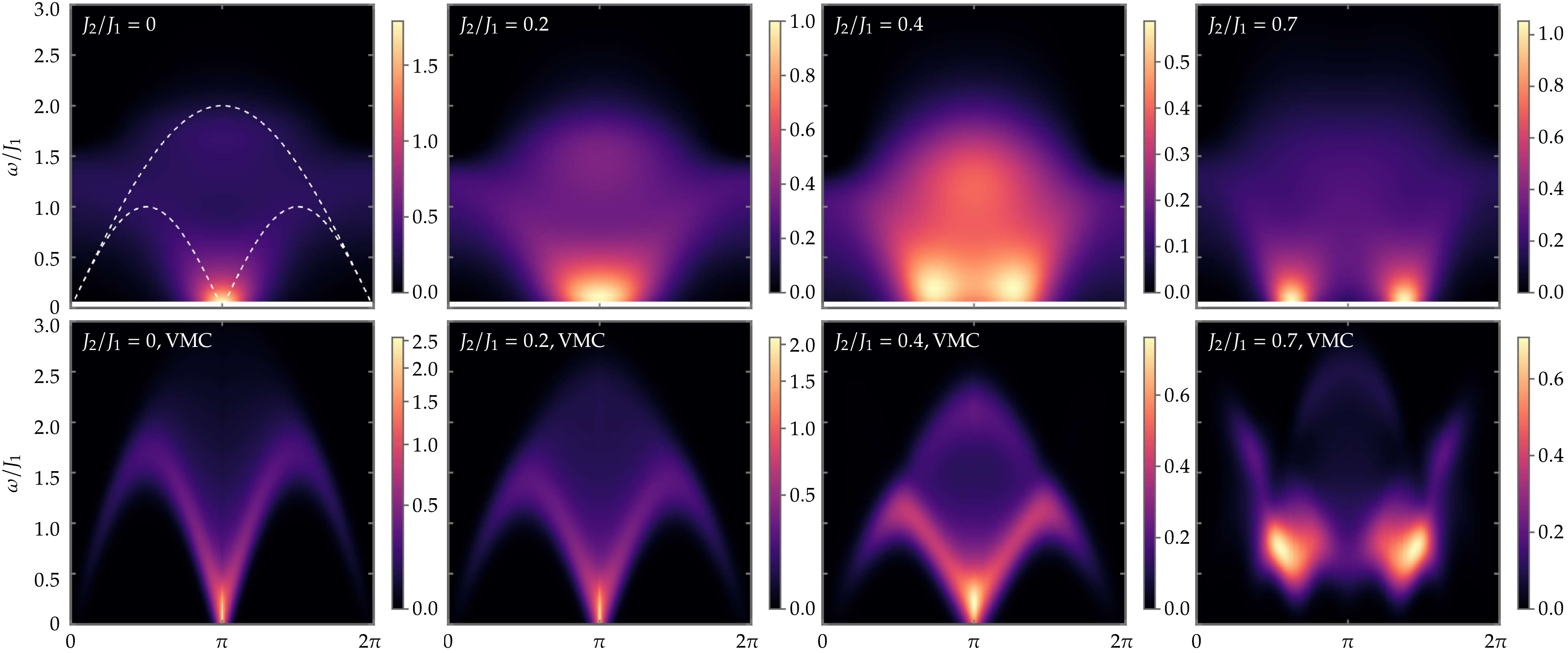}
		\caption{(upper line) Dynamical spin structure factor for the one-dimensional spin chain with $N=10$ at $T=0.1J_1$ from Keldysh PFFRG. For $J_2/J_1=0$ we can see a strong peak at $q=\pi$. The spinon continuum is visible above the main excitation. A comparison with the two-spinon energy bounds is shown by white dashed lines. Since the Keldysh PFFRG quantitatively does not match the frequency scale we also fitted the pre-factor to compare with our results. For $J_2/J_1>J_\text{KT}$ we should see a gapped spectrum, which is not visible in our results. For larger $J_2/J_1$ we get a splitting of the main peak into two peaks. (lower line) VMC data reproduced from Ref. \cite{BeccaSpinChain} with an increased broadening to reflect the finite temperature of our method as discussed in the Supplemental Material \cite{NoteSM}.
		}
		\label{fig:SpinChainFull}
	\end{figure*}

	From the zero-dimensional dimer we now shift focus to the one-dimensional spin chain with nearest ($J_1$) and next-nearest ($J_2$) neighbor AFM interactions.
	
	\subsubsection{RPA Results}\label{sec:SpinChainRPA}
	Similar to the AFM dimer we first discuss the results in the classical $S\to\infty$ limit. The RPA results -- coinciding with the FRG -- are shown in Fig.~\ref{fig:RPASpinChain}. Excitations are observable only for $\omega\approx0$ with a maximum at $q=\pi$. However, we would expect excitation branches for all $q$ with a notable dispersion in $\omega$, since the classical $S\to\infty$ limit should be well described by linear spin wave theory (SWT) \cite{spinWave1,spinWave2}. There, in first order, the dispersion is given by
	\begin{align}
		\omega\propto J\abs{\sin(q)},\quad\text{with } q\in\{0,2\pi\}.
	\end{align}
	
	However, this holds only under the assumption of a response around an ordered ground state, which is in contrast with the FRG 
	where one assumes the initial state to be paramagnetic and thus maximally disordered~\cite{ReutherFirstFRG}. As outlined above, achieving magnetic order in the FRG would require the presence of a finite order parameter within the flow. Since, we exclude any magnetic order parameter, the flow diverges in the vicinity of any ordering phase, confining us to the paramagnetic regime.
	This limitation may also explain why small temperatures cannot be reached in the RPA calculations since the $t$-channel is known to enhance the ordering tendencies, while the -- in the RPA vanishing -- $s$ and $u$ channel suppress ordering~\cite{DissTobi,DissReuther}. Without a flowing order parameter, the divergence reflects the onset of ordering. This observation also sets the stage for the remainder of this study, as systems well-described by SWT may lack certain features in FRG, especially around $q=0$. 
	As mentioned above, to permit a flow into the symmetry broken phase, the Keldysh PFFRG would have to allow for a magnetic field, which induces a spin rotational symmetry breaking. 
	
	\subsubsection{FRG Results}
	Following the RPA discussion we now focus on the full FRG solution and analyze how the additional channels and the interplay between them enhances the results to be able to describe the full quantum $S=1/2$ Heisenberg model.
	
	When analyzing this model one normally expects the typical excitations to carry spin $S=1$ which are the so-called magnons. For spin liquids however, it is possible for the system to exhibit $S=1/2$ spinons as the principal excitation, which implies a fractionalization of the spin quantum number~\cite{Fractionalization}. In the case of the spin chain these can be identified with domain walls, which can propagate freely through the chain once they are created until they are annihilated again. Thus, calculating the spin structure factor for the Heisenberg spin chain is an excellent test of our method to ascertain whether emergent excitations can be resolved. Furthermore, the $J_1$-$J_2$ phase diagram is already well studied and can be used as a benchmark~\cite{SpinChainGap}. Specifically, for $J_2=0$ we have the isotropic spin chain, which is exactly solvable using the Bethe Ansatz~\cite{Bethe}. However, a closed expression for the dynamical spin structure factor does not exist. (This is only the case for a spin chain with inverse-square superexchange \cite{SpinChainInverseSquare}.)  At the Kosterlitz-Thouless point $J_2/J_1=J_\text{KT}=0.24116(7)$~\cite{KTTransition} the system exhibits a phase transition from gapless to gapped \cite{SpinChainGap}.
	
	For $J_2/J_1=0.5$ the system is known as the Majumdar-Ghosh model~\cite{majumdarGhosh} which, akin to the $J_2=0$ case is analytically solvable. Lastly, for $J_2\to\infty$ the system transitions into two decoupled spin chains, which closes the gap again. The results for the intermediate values can be compared against various numerical techniques as the variational Monte Carlo approach~\cite{BeccaSpinChain, SpinChainFiniteT} and also experimental data~\cite{SpinChainMeasurement}. 
	
	We calculate the dynamical spin structure factor for correlations up to $l=10$ sites with different $J_2/J_1$ at $T=0.1J_1$. The temperature was chosen as a compromise between $T\ll J_1$ and the effects described in Sec.~\ref{sec:dimer}. The resulting spectra are shown in Fig.~\ref{fig:SpinChainFull}. Since, $\Im(\chi(\omega=0))=0$ is given by symmetry we only show the values for $\omega>0$, as the limit $\omega\to0$ in the structure factor Eq.~\eqref{eq:structreFactor} is ill-defined. The smallest frequency we use is $\omega=0.06$. This choice is arbitrary and results from the susceptibility calculation discretization. For larger $\omega$, some susceptibilities are slightly smaller than zero. We attribute this to numerical inaccuracies and set those values to zero since they are at least two orders of magnitude smaller compared with the results for small frequencies. Furthermore, we reproduced and added the corresponding numerical variational Monte Carlo (VMC) results from Ref. \cite{BeccaSpinChain} as a comparison for each coupling ratio $J_2/J_1$.

        \begin{figure*}[t!]
		\includegraphics[width=.33\linewidth]{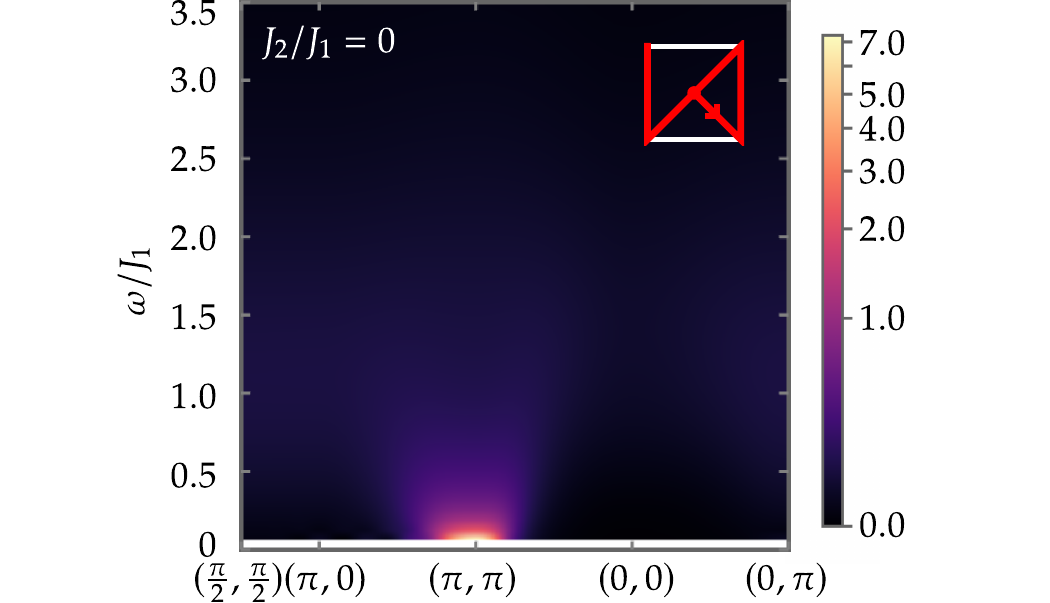}
		\includegraphics[width=.33\linewidth]{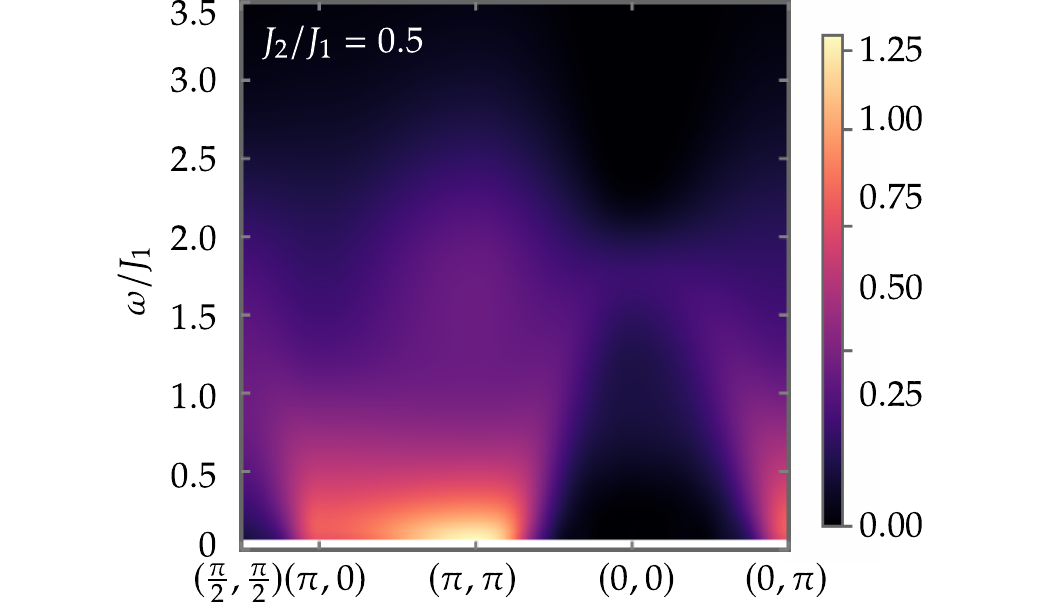}
        \includegraphics[width=.33\linewidth]{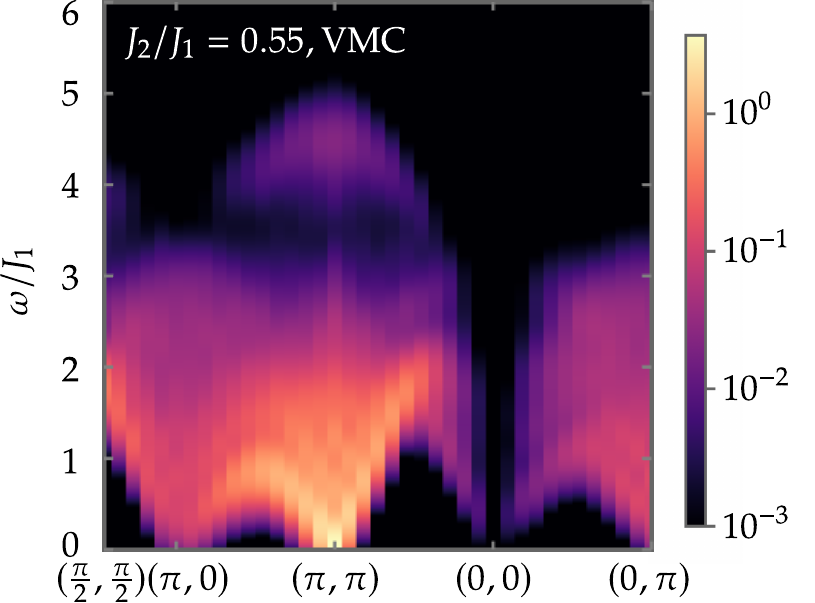}
		\label{fig:SquareFrustrated}
		\caption{Dynamical spin structure factor for the square lattice at $T=0.1J_1$ with a maximum distance of $l=7$ lattice sites along the depicted high symmetry line. $J_2=0$ is the unfrustrated case and is taken at $\Lambda=0.48J_1$, as the flow had to be stopped due to the ordering tendencies. The sharp peak at $\bm{q}=(\pi,\pi)$ signals the antiferromagnetic ordering. Due to the large cutoff scale sharp branches are not visible. For the frustrated case with $J_2/J_1=0.5$ we see a strong signal also at $\bm{q}=(\pi,0)$ which is to be expected when comparing with the VMC results reproduced from Ref. \cite{VMCSquareLattice} as discussed in the Supplemental Material \cite{NoteSM}. For both $J_2=0$ and $J_2=0.5J_1$ the branch to $\bm{q}=(0,0)$ is not resolved in the Keldysh formalism.
		}
		\label{fig:Square}
	\end{figure*}
	
	All structure factors from the Keldysh PFFRG show the typical dome structure as well as the spinon continuum. Below the Majumdar-Ghosh point we get one maximum at $q=\pi$. In this phase one would also expect branches going down to $q=0$ as discussed in Sec.~\ref{sec:SpinChainRPA} and visible in the VMC data. However, this is not resolved in our analysis as the susceptibility continuously decreases away from $q=\pi$. This means, that the additional channels cannot remedy the fact, that the paramagnetic phase which we cannot leave during the flow, suppresses these kind of branches. 

    Additionally, the pseudo-fermion number constraint, is implemented only on average and thus does not prevent the inclusion of doubly occupied or empty sites in the flow. This, in turn, introduces finite-size spin chain effects into the results. Consequently, this is likely the primary cause of the broad continuum observed between $\omega = J$ and $\omega = 1.5J$, as demonstrated by the comparisons in the Supplemental Material \cite{NoteSM}.
	
	Furthermore, a differentiation between gapless and gapped spectra is not possible because only the $J_2=0$ spectrum should be gapless. For $J_2/J_1\leq0.5$, our inability to resolve the gap could be attributed to its relatively small sizes and the finite temperatures in our calculations but for $J_2/J_1=0.75$ the gap is $\Delta/J_1\approx0.4$ \cite{SpinChainGap}, which should be well resolved. Thus we can argue, that the Keldysh-PFFRG is not able to resolve the difference between a gapped and a gapless spectrum. This is again due to the nature of FRG, as we need a finite mass term for the excitations to create a gap in the spectrum. This mass term however is connected to the order parameter of an ordered phase and would need to be specifically included into the flow to become finite. Otherwise it is just neglected, which is why our implementation is not able to differentiate between gapped and gapless spectra. 
	
	At $J_2=0.5 J_1$, the exactly-solvable Majumdar-Ghosh point, we find incommensurate correlations for $\omega \to 0$. Analytically, the dominant correlations should still reside at $q=\pi$ and only for larger $J_2$ move toward $q=\pi/2$. This finding, however, is consistent with Matsubara PFFRG calculations using the \texttt{PFFRGSolver.jl} software~\cite{Multiloop,PFFRGSolver}, where we find the exact same values of $q$ for the dominant correlations.  For $J_2/J_1>0.5$ we, however, see the correct behavior of the system, which means a continuous movement of the peak position to $q=\pi/2$. Also, the peak intensity of the susceptibility at $J_2/J_1=0.75$ shows the trend of transforming into two decoupled spin chains as found for $J_2 \to \infty$.
	
	This means, that PFFRG is able to qualitatively resemble the structure of the features in the susceptibility and improve significantly beyond the results of the RPA calculation, which was merely able to reflect the paramagnetic properties of the initial state. Nevertheless, the inability of our method to enter a symmetry broken phase still prohibits some features as well as a proper quantitative analysis for the FRG.
    
    \begin{figure*}[t!]
		\includegraphics[width=.33\linewidth]{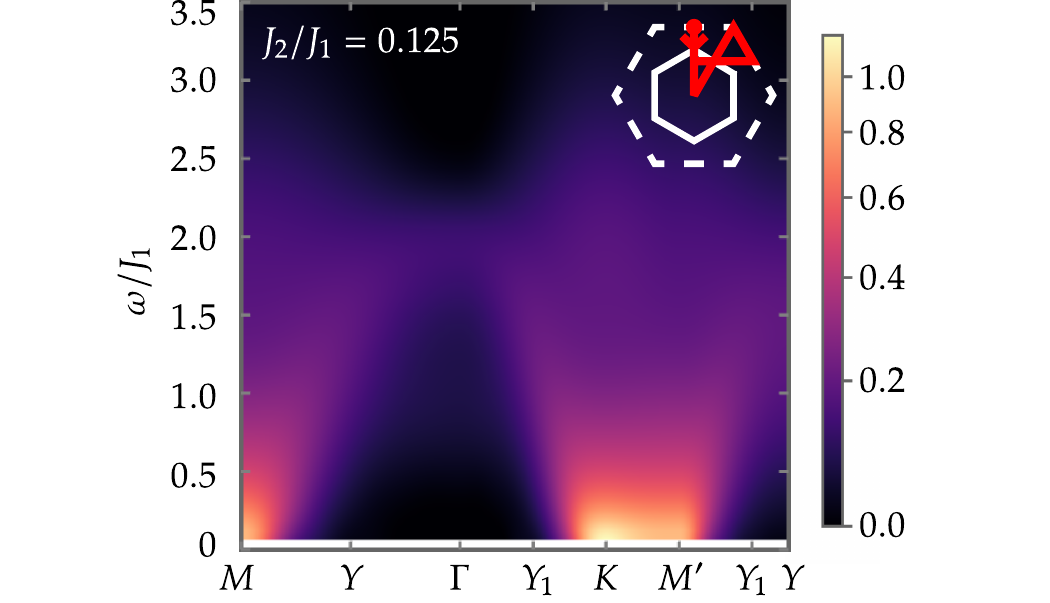}
		\includegraphics[width=.33\linewidth]{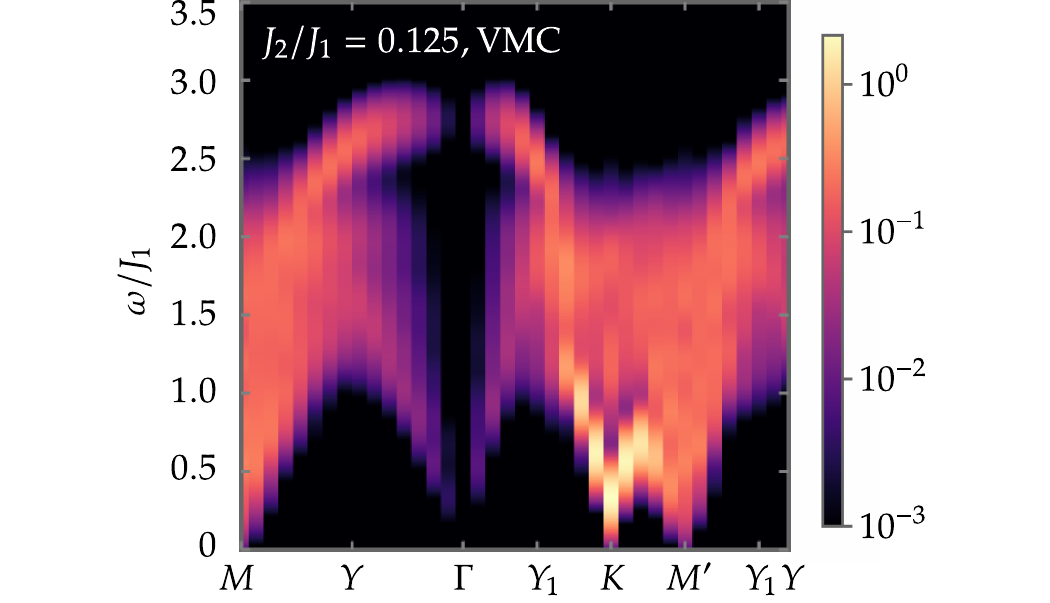}
        \includegraphics[width=.33\linewidth]{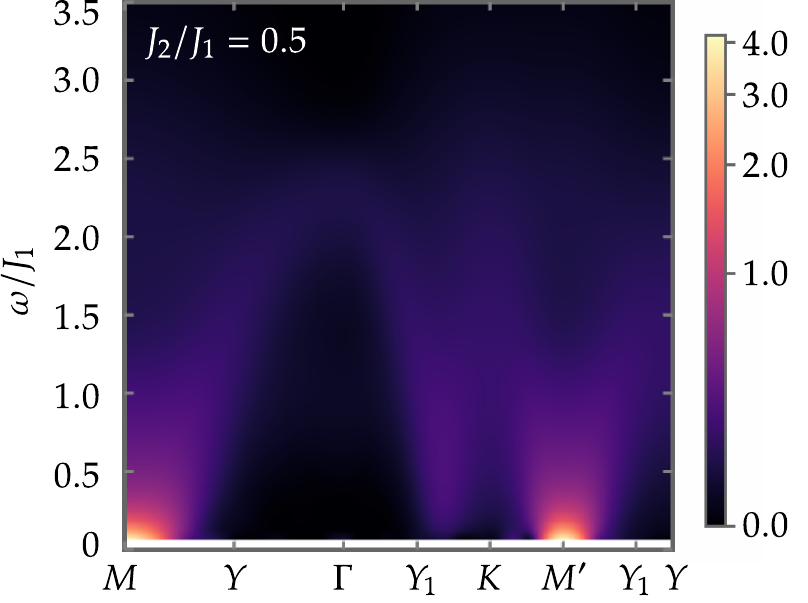}
		\caption{Dynamical spin structure factor for the $J_1-J_2$ triangular lattice. For $J_2=0.125J_1$, which is inside the spin liquid phase, we compare the Keldysh PFFRG results to the VMC results reproduced from Ref. \cite{Ferrari-2019}. Similar to the bipartite spin chain and square lattice, the branch to $q=\Gamma$ is not resolved and the absolute frequency scale does not match. We can nonetheless resolve the main branches as well as the peak at $K$, which only appears in the VMC results after Gutzwiller projection~\cite{Ferrari-2019}. For $J_2/J_1=0.5$ we see a tendency to stripe ordering due to the large weight at $\omega\approx0$, but the finite temperature still allows the Keldysh PFFRG to be applied. 
		}
		\label{fig:Triangular}
	\end{figure*}
	
	\subsection{(Frustrated) Square Lattice}\label{sec:square}
	
	Having discussed zero- and one-dimensional system, we now turn our attention to the two-dimensional square lattice. For only nearest-neighbor antiferromagnetic interactions the system is known to have a N\'eel-ordered ground state with ordering vector $\bm{q}=(\pi,\pi)$ at $T=0$. The elementary excitations can be described by spin wave theory and are gapless. In first order SWT the dispersion is given by
	\begin{align}
		\omega\propto J\sqrt{1-\gamma(\bm{q})^2},\quad\text{with } 2\gamma(\bm{q})=\cos(q_x)+\cos(q_y).
	\end{align}
	Note, that there are additional modulations when computing the spin susceptibility, which suppress the intensity around $q=(0,0)$. While higher-order corrections to this result can be computed~\cite{spinwaveCorrections}, it is important to note that spin wave theory, being a semiclassical approach, fails to account for certain deviations, particularly a suppression of spectral weight at $\bm{q} = (\pi, 0)$~\cite{SquareLatticeDipExp}. These deviations, however, are captured by a series expansion approach~\cite{squareLatticeSeriesExpansion,squareLatticeSeriesExpansionCorrection}.
	
	We analyze the square lattice with a maximal correlation length of $l=7$ for both the pure nearest-neighbor antiferromagnet and a parametrically frustrated parameter point with next-nearest neighbor coupling $J_2=0.5J_1$. In the pure nearest-neighbor model ($J_2=0$) we encounter a divergence in the flow signaling an ordered ground state. This should in principle not be the case because the Mermin-Wagner theorem prohibits breaking of the continuous spin-rotation symmetry in two dimensions for $T>0$. 
	The PFFRG however, especially at low temperatures, shows enhanced ordering tendencies, which result in the divergence as already discussed in Sec.~\ref{sec:dimer}. To avoid this, we could have performed the calculations at a higher temperature, where there would be no divergence in the flow (i.e. $T\approx J$). This approach, though, would significantly smear out all signatures in the dynamic structure factor. Thus, we stay at $T=0.1J$ considering that the $J_2=0$ case should only be taken as a reference for the frustrated parameters. The resulting spin structure factors are shown in Fig.~\ref{fig:Square}. When comparing these results to VMC~\cite{VMCSquareLattice} or other numerical approaches~\cite{SquareLatticeMCStructureFactor,Piazza-2015,Shao-2017,Yu-2018,Hu-2013} we observe a qualitatively good agreement, except for the branches leading to $\bm{q} = (0, 0)$ -- an issue also encountered in the one-dimensional spin chain. This we again attribute to the paramagnetic nature of the FRG flow. Nonetheless, within the frustrated regime, the PFFRG successfully captures the gapless excitations at $(\pi,0)$ and $(0,\pi)$ for $J_{2}/J_{1}=0.5$, which can be ascribed to the gapless $\mathbb{Z}_{2}$ Dirac spin liquid ground state (the Z2Azz13 state of Ref.~\cite{Wen-2002}) in this region, which features four Dirac cones at $(\pm\frac{\pi}{2},\pm\frac{\pi}{2})$ in the spinon spectrum resulting in gapless two-spinon excitations at $(\pi,0)$ and $(0,\pi)$ in addition to those at $(0,0)$ and $(\pi,\pi)$~\cite{VMCSquareLattice,Hu-2013}. The missing arc between $\bm{q}=(\pi,0)$ and $(\pi,\pi)$ in the frustrated case could be due to the one-loop implementation of the PFFRG but also because of the relatively strong broadening due to the temperature. Nevertheless, the comparison between the frustrated and the non-frustrated model shows a clear formation of an excitation continuum in the frustrated case, while the excitations remain relatively localized for $J_2=0$. There however, clear branches are absent due to the divergence of the flow, leading to the evaluation of the susceptibility at finite $\Lambda$, which introduces broadening and obscures any clear dispersion or even formation of spinon branches.
	
	Most importantly, in the spin liquid regime, we are able to resolve the spinon continuum, and successfully describe the development of gapless excitations at $(0,\pi)$ and $(\pi,0)$ characterizing the nature of the spin liquid, while the missing branches out of $\bm{q}=(0,0)$ and between $\bm{q}=(\pi,0)$ and $\bm{q}=(\pi,\pi)$ are most likely not captured due to finite temperature effects. Thus, we can support the claim, that higher dimensional systems are more favorable to the FRG formalism, as approaching the critical mean-field dimension $n=4$ reduces the effect of non-renormalizable large local couplings~\cite{DMF2RG}. 
    
\subsection{Triangular Lattice}
With all the prior benchmarks we can start using the Kelydsh PFFRG on models which are not straightforwardly amenable to standard techniques. The frustrated $J_1-J_2$ triangular lattice with the corresponding spin liquid phase has recently garnered much attention, with a particular focus on its dynamical signatures. With classical spins and $J_2=0$, the triangular lattice minimizes its energy by forming the so-called $120^\circ$ order. While replacing the classical spins with quantum spins was proposed as a route by Anderson to potentially realize a resonating valence bond liquid \cite{TriangularAnderson}, it was later shown numerically, that for $J_2\approx0.07J_1$ the $120\degree$ order melts and forms a spin liquid \cite{Triangular1,Triangular2,TriangularYasir}. While there exists no experimental confirmation on the nature of the spin liquid, there are arguments for a gapped \cite{Triangular1,Triangular2} and a gapless spin liquid \cite{TriangularYasir,Hu-2019}. Since the Keldysh formalism is not able to resolve gaps, we refrain from commenting on this issue. We can however analyze how the Keldysh PFFRG results compare with the available results for the dynamical structure factor from VMC~\cite{Ferrari-2019} similar to the spin chain and square lattice. This is shown in Fig.~\ref{fig:Triangular} for $J_2=0.125J_1$ which is inside of the spin liquid phase. We can see that both the Keldysh PFFRG display a dominant peak at $K$ and subdominant peak at $M'$ as observed in VMC~\cite{Ferrari-2019}, random phase approximation~\cite{Willsher-2025} and density matrix renormalization group studies~\cite{Sherman-2023,Drescher-2023}. While they are gapped in the VMC data, this gap shrinks with the increased broadening parameter $\eta=0.1$ used for the calculation, mimicking the effects the temperature might have on our results. Furthermore, as discussed in Ref.~\cite{Ferrari-2019}, the branch at $K$ only becomes visible after applying a Gutzwiller projection, which enforces the single-occupancy constraint, i.e., accounts for the fluctuations of the temporal component of the gauge field. This means, that despite having the pseudo-fermion constraint violated, we are still able to capture the correct physical effects of spinon interactions (which are a beyond mean-field effect), which could be strong within the picture of a U(1) Dirac spin liquid ground state. These lead to the appearance of a two-spinon bound state with a minimum at $K$ which has been argued to be reflective of the incipient $120\degree$ order~\cite{Willsher-2025}, however, triplet monopoles yield a similar picture~\cite{Budaraju-2025}. We have also calculated the dynamical spin structure factor for $J_2=0.5J_1$ which lies outside of the spin-liquid phase. In principle, this means that the flow should diverge in the same way as for the square lattice. However, the geometric frustration of the triangular lattice together with the temperature of $\beta=0.1J$ allows us to stay in the paramagnetic regime during the flow, and therefore make a prediction for the dynamics of the corresponding phase. We see in Fig.~\ref{fig:Triangular}, that the $K$-point weight, which for $J_2<0.08J_1$ dominated the low energy structure factor, now vanishes, being replaced by the $M$-point ordering. Further, the branches are much sharper, which we attribute to the incipient stripe ordered phase, which is solely suppressed by the finite temperature. In a similar manner we were able to calculate the $J_2<0.07J_1$ data, which is shown in the supplemental material \cite{NoteSM} and hosts the expected $K$-ordering.
	\subsection{Honeycomb-Kitaev Model}\label{sec:kitaev}

    	\begin{figure*}[t!]
		\includegraphics[width=.49\linewidth]{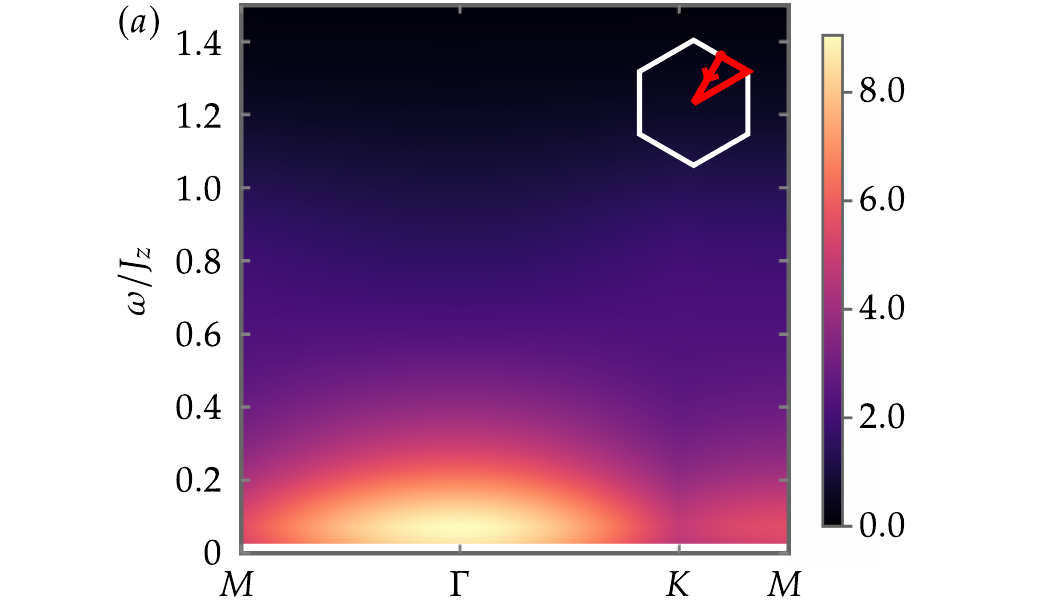}
		\includegraphics[width=.49\linewidth]{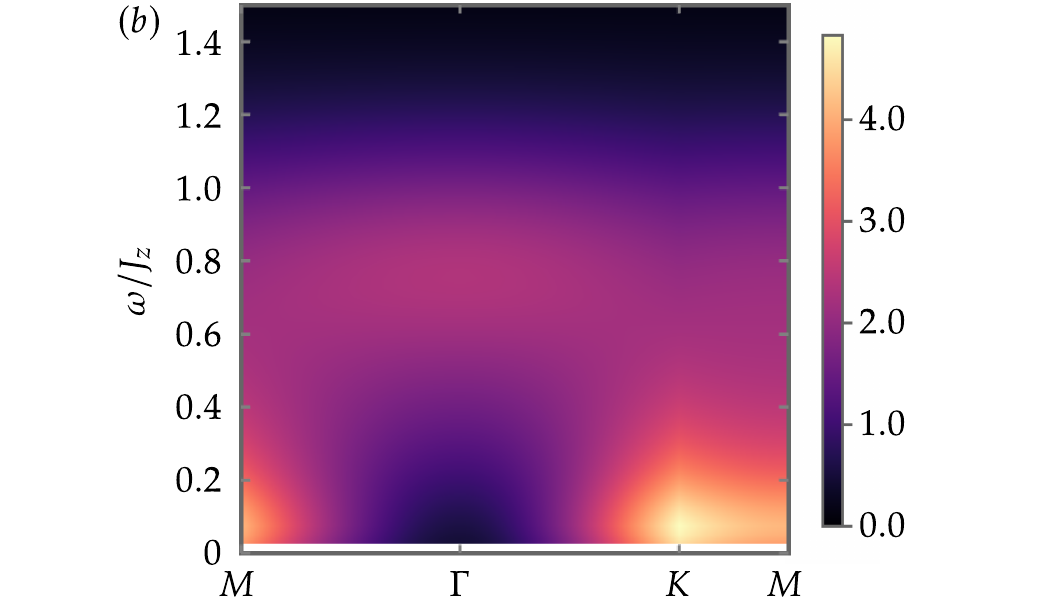}
		\label{fig:KitaevAFM}
		\caption{Dynamical spin structure factor for the isotropic Honeycomb-Kitaev model at $\beta=10/J$. The (a) ferromagnetic and the (b) anti-ferromagnetic case show the typical excitation structure of the Kitaev model. The high symmetry path is chosen to match Ref.~\cite{KitaevStructureFactor} and is depicted in panel (a). The frequency scale does not match the exact results and the two-flux gap is not resolved.
		}
		\label{fig:Kitaev}
	\end{figure*}
	
	Going beyond the isotropic Heisenberg model, we introduce directional couplings $J^{xx},J^{yy}$ and $J^{zz}$ in spin space, which allow us to calculate the dynamical spin structure factor of a much larger class of magnetic systems. We chose the Honeycomb-Kitaev model as a benchmark, as it is analytically solvable via the introduction of Majorana fermions \cite{KitaevStructureFactor2, KitaevStructureFactor}. The Kitaev-Honeycomb model is defined on a typical honeycomb lattice with interactions chosen such that only one of the $J^{aa}$ is non-zero along a specific bond direction~\cite{KitaevReview}.
	
	Furthermore, the Kitaev-Honeycomb model suits the PFFRG as one can show that only nearest-neighbor interactions play a role in the flow and we thus can calculate the dynamics with very small numerical effort. The dynamical spin structure factors for the isotropic ferromagnetic and anti-ferromagnetic interactions are depicted in Fig.~\ref{fig:Kitaev}. Both resemble the structure of the exact solution given in Ref.~\cite{KitaevStructureFactor}. In the ferromagnetic case, we find a broad spectrum at low frequency, which is lightly peaked around the $\Gamma$ point with a continuum of excitations which falls of quickly for larger frequencies. In the antiferromagnetic case, the structure factor is peaked at the K point at small frequencies and shows vanishing spectral weight in a region around $\Gamma$. At higher energies, this spectral gap is arched over by a continuum in the structure factor, as found in the analytical solution.
	However, in both cases, the two-flux gap is not visible. This again is due to the fact, that the FRG is not able to resolve gapped spectra without the inclusion of an additional flowing order parameter. Furthermore, the energy scale of the PFFRG results is approximately five times smaller than the corresponding exact results which might also be a result of the missing order parameter or is due to the paramagnetic starting point of our technique. This benchmark shows, that our approach is only able to give correct qualitative results, but quantitatively lacks in comparability. 
	
	\section{Conclusion}\label{sec:conclusion}
    
	The Keldysh formulation of the pseudo-fermion functional renormalization group (FRG) stands out as a novel many-body framework capable of directly evaluating dynamic spin structure factors of spin models in an unbiased way. We have benchmarked our technique for low-dimensional spin models with diagonal interactions such as the zero-dimensional dimer, the one-dimensional spin chain, and the two-dimensional square and triangular lattice (also with frustrated interactions) Heisenberg models. Furthermore, we assessed the dynamical spin structure factor for the canonical Kitaev-Honeycomb model for which exact results are available. As expected, the Keldysh PFFRG results significantly improve upon those from RPA and seem particularly reliable for systems with an absence of conventional magnetic order. It would constitute an important future endeavor to allow for inclusion of Zeeman pinning fields in the Keldsyh formulation of PFFRG which would permit a flow into the ordered phase without encountering unphysical divergencies in the flow down till zero cutoff, thus enabling us to faithfully capture magnon branches. Therefore, the presently developed Keldysh PFFRG is best suited for models in the highly frustrated regime displaying putative spin liquid behavior close to the paramagnetic regime, which is indeed the regime of interest where most pressing questions at the interface with quantum materials lie. While factors such as pseudo-fermion number violation, the paramagnetic initial state and finite-temperature effects limit the quantitative accuracy of the Keldysh PFFRG results, we believe that this work opens pathways to further refinements which mitigate these factors, and will thus potentially make it a sharpened tool for exploring the intricate excitation spectra of models and materials. These pathways include - as mentioned in the main text - employing the pseudo-Majorana decomposition \cite{PseudoMajorana1,PseudoMajorana2}, incorporating the Popov-Fedotov technique, and extending the flow to account for non-zero magnetic fields \cite{PFFRGMagnetic,PMFRG_Magnetic}. While the former will definitely improve the results of the RG flow by operating within the physical Hilbert space of the spin model, the latter might allow to resolve the results with magnetic ordering and therefore could lead to an enhancement of the expected branchs around ${\mathbf q}=0$. For any gap to be visible one probably has to introduce a two-step process with an additional order parameter to describe the gap, which the standard FRG is not able to do.
    
    Improving upon the numerical performance should lead to the possibility of mapping phase diagrams instead of singular points in the parameter space of Hamiltonian couplings. Furthermore, employing the full Keldysh formalism \cite{Nonequilibrium,Nonequilbrium2} will allow for accessing non-equilibrium phenomena in frustrated magnets such as computation of spinon currents and transport coefficients, which remain {\it terra incognita}. Given the arrival of a plethora of three-dimensional candidate quantum spin liquid materials based on novel lattices of spin-$1/2$ and $1$ magnetic moments for which neutron scattering profiles are currently getting available, the Keldysh PFFRG approach would serve as a valuable tool in identifying the precise nature of the spin liquids by making connections to the underlying microscopic model if further improved on the current points or used with the current shortcomings in mind. 
	

	
	\section{Acknowledgements}
	We thank Francesco Ferrari for providing us his VMC data from Ref.~\cite{Ferrari-2019,BeccaSpinChain,VMCSquareLattice} as the main benchmark in this work. We thank Nils Niggemann and Johannes Reuther for fruitful discussions. We also want to thank Anxiang Ge and Nepomuk Ritz for benchmarking the Keldysh implementation against the SIAM code. Further we want to explicitly thank Benedikt Schneider for highlighting a problem in the susceptibility calculation. 
	The work is funded by the Deutsche Forschungsgemeinschaft (DFG, German Research Foundation) through Project-ID 258499086 - SFB 1170, through the W\"urzburg-Dresden Cluster of Excellence on Complexity and Topology in Quantum Matter -- \textit{ct.qmat} Project-ID 390858490 - EXC 2147 and the Research Unit QUAST by the DFG Project-ID 449872909 - FOR5249. The work of Y.I. was performed in part at the Aspen Center for Physics, which is supported by National Science Foundation Grant No.~PHY-2210452 and a grant from the Simons Foundation (1161654, Troyer). This research was supported in part by grant NSF PHY-2309135 to the Kavli Institute for Theoretical Physics (KITP). Y.I. acknowledges support from the ICTP through the Associates Programme and from the Simons Foundation through Grant No.~284558FY19.
	
	
	\bibliography{KPFFRG.bib}
    \newpage
    \appendix
    \begin{widetext}
    \section{Full FRG Flow equations}\label{app:flowEquations}
In this section we present the flow equations for the Kitaev model, which can be easily simplified into the pure SU(2) symmetric Heisenberg model by $\Gamma_{xx}=\Gamma_{yy}=\Gamma_{zz}=\Gamma_s$. The flow equations are obtained by inserting the vertex decomposition Eq.~\eqref{eq:vertexDecomp} into the flow Eq.~\eqref{eq:flowEq}. After performing all spin sums we arrive at the full flow equations.
\begin{align}
	\frac{\mathrm{d}}{\mathrm{d}\Lambda}\xvortex{1}{2}{1}{2}{1'}{2'}{1}{2}=\frac{1}{2\pi}&\sum_{\alpha_{3^\noprime}\alpha_{3'}}\sum_{\alpha_{4^\noprime}\alpha_{4'}}\int\mathrm{d}\omega_3\int\mathrm{d}\omega_4\left(G^\Lambda_{\alpha_3\alpha_{3'}}(\omega_3)S^\Lambda_{\alpha_4\alpha_{4'}}(\omega_4)+G^\Lambda_{\alpha_4\alpha_{4'}}(\omega_4)S^\Lambda_{\alpha_3\alpha_{3'}}(\omega_3)\right)\nonumber\\\nonumber
	\bigg\{\big[\phantom+&\xvortex{1}{2}{3}{4}{1'}{2'}{1}{2} \dvortex{3}{4}{1}{2}{3}{4}{1}{2}-\yvortex{1}{2}{3}{4}{1'}{2'}{1}{2}\zvortex{3}{4}{1}{2}{3}{4}{1}{2}\\\nonumber
	-&\zvortex{1}{2}{3}{4}{1'}{2'}{1}{2}\yvortex{3}{4}{1}{2}{3}{4}{1}{2}+\dvortex{1}{2}{3}{4}{1'}{2'}{1}{2}\xvortex{3}{4}{1}{2}{3}{4}{1}{2}\big]\\\nonumber
	\cross&\dird{1'}{2'}{3}{4}\dird{3}{4}{1}{2}
	\\\nonumber
	-2\sum_{i_3}&\xvortex{1}{4}{1}{3}{1'}{4}{1}{3}\xvortex{3}{2}{4}{2}{3}{2'}{3}{1}\dird{1'}{4}{1}{3}\dird{3}{2}{4}{2}
	\\\nonumber
	+\big[&\xvortex{1}{4}{1}{3}{1'}{4}{1}{2}\xvortex{3}{2}{2}{4}{3}{2'}{2}{2}-\xvortex{1}{4}{1}{3}{1'}{4}{1}{2}\yvortex{3}{2}{2}{4}{3}{2'}{2}{2}\\\nonumber
	-&\xvortex{1}{4}{1}{3}{1'}{4}{1}{2}\zvortex{3}{2}{2}{4}{3}{2'}{2}{2}+ \xvortex{1}{4}{1}{3}{1'}{4}{1}{2}\dvortex{3}{2}{2}{4}{3}{2'}{2}{2}\big]\\\nonumber
	\cross&\dird{1'}{4}{1}{3}\dird{3}{2'}{2}{4}
	\\\nonumber
	+\big[&\xvortex{1}{4}{3}{1}{1'}{4}{1}{1}\xvortex{3}{2}{4}{2}{3}{2'}{1}{2}-\yvortex{1}{4}{3}{1}{1'}{4}{1}{1}\xvortex{3}{2}{4}{2}{3}{2'}{1}{2}\\\nonumber
	-&\zvortex{1}{4}{3}{1}{1'}{4}{1}{1}\xvortex{3}{2}{4}{2}{3}{2'}{1}{2}+\dvortex{1}{4}{3}{1}{1'}{4}{1}{1}\xvortex{3}{2}{4}{2}{3}{2'}{1}{2}\big]\\\nonumber
	\cross&\dird{1'}{4}{3}{1}\dird{3}{2'}{4}{2}
	\\\nonumber
	+\big[&\xvortex{3}{2}{1}{4}{3}{2'}{1}{2}\dvortex{1}{4}{3}{2}{1'}{4}{1}{2}+\yvortex{3}{2}{1}{4}{3}{2'}{1}{2}\zvortex{1}{4}{3}{2}{1'}{4}{1}{2}\\\nonumber
	+&\zvortex{3}{2}{1}{4}{3}{2'}{1}{2}\yvortex{1}{4}{3}{2}{1'}{4}{1}{2}+\dvortex{3}{2}{1}{4}{3}{2'}{1}{2}\xvortex{1}{4}{3}{2}{1'}{4}{1}{2}\big]\\
	\cross&\dird{3}{2'}{1}{4}\dird{1'}{4}{3}{2}\bigg\}
\end{align}

\begin{align}
	\frac{\mathrm{d}}{\mathrm{d}\Lambda}\yvortex{1}{2}{1}{2}{1'}{2'}{1}{2}=\frac{1}{2\pi}&\sum_{\alpha_{3^\noprime}\alpha_{3'}}\sum_{\alpha_{4^\noprime}\alpha_{4'}}\int\mathrm{d}\omega_3\int\mathrm{d}\omega_4\left(G^\Lambda_{\alpha_3\alpha_{3'}}(\omega_3)S^\Lambda_{\alpha_4\alpha_{4'}}(\omega_4)+G^\Lambda_{\alpha_4\alpha_{4'}}(\omega_4)S^\Lambda_{\alpha_3\alpha_{3'}}(\omega_3)\right)\nonumber\\\nonumber
	\bigg\{\big[-&\xvortex{1}{2}{3}{4}{1'}{2'}{1}{2}\zvortex{3}{4}{1}{2}{3}{4}{1}{2}+\yvortex{1}{2}{3}{4}{1'}{2'}{1}{2} \dvortex{3}{4}{1}{2}{3}{4}{1}{2}\\\nonumber
	-&\zvortex{1}{2}{3}{4}{1'}{2'}{1}{2}\xvortex{3}{4}{1}{2}{3}{4}{1}{2}+\dvortex{1}{2}{3}{4}{1'}{2'}{1}{2}\yvortex{3}{4}{1}{2}{3}{4}{1}{2}\big]\\\nonumber
	\cross&\dird{1'}{2'}{3}{4}\dird{3}{4}{1}{2}
	\\\nonumber
	-2\sum_{i_3}&\yvortex{1}{4}{1}{3}{1'}{4}{1}{3} \yvortex{3}{2}{4}{2}{3}{2'}{3}{1}\dird{1'}{4}{1}{3}\dird{3}{2}{4}{2}
	\\\nonumber
	+\big[-& \yvortex{1}{4}{1}{3}{1'}{4}{1}{2}\xvortex{3}{2}{2}{4}{3}{2'}{2}{2}+ \yvortex{1}{4}{1}{3}{1'}{4}{1}{2}\yvortex{3}{2}{2}{4}{3}{2'}{2}{2}\\\nonumber
	-&\yvortex{1}{4}{1}{3}{1'}{4}{1}{2}\zvortex{3}{2}{2}{4}{3}{2'}{2}{2}+ \yvortex{1}{4}{1}{3}{1'}{4}{1}{2}\dvortex{3}{2}{2}{4}{3}{2'}{2}{2}\big]\\\nonumber
	\cross&\dird{1'}{4}{1}{3}\dird{3}{2'}{2}{4}
	\\\nonumber
	+\big[-&\xvortex{1}{4}{3}{1}{1'}{4}{1}{1}\yvortex{3}{2}{4}{2}{3}{2'}{1}{2}+\yvortex{1}{4}{3}{1}{1'}{4}{1}{1}\yvortex{3}{2}{4}{2}{3}{2'}{1}{2}\\\nonumber
	-&\zvortex{1}{4}{3}{1}{1'}{4}{1}{1}\yvortex{3}{2}{4}{2}{3}{2'}{1}{2}+\dvortex{1}{4}{3}{1}{1'}{4}{1}{1}\yvortex{3}{2}{4}{2}{3}{2'}{1}{2}\big]\\\nonumber
	\cross&\dird{1'}{4}{3}{1}\dird{3}{2'}{4}{2}
	\\\nonumber
	+\big[&\xvortex{3}{2}{1}{4}{3}{2'}{1}{2}\zvortex{1}{4}{3}{2}{1'}{4}{1}{2}+\yvortex{3}{2}{1}{4}{3}{2'}{1}{2}\dvortex{1}{4}{3}{2}{1'}{4}{1}{2}\\\nonumber
	+&\zvortex{3}{2}{1}{4}{3}{2'}{1}{2}\xvortex{1}{4}{3}{2}{1'}{4}{1}{2}+\dvortex{3}{2}{1}{4}{3}{2'}{1}{2}\yvortex{1}{4}{3}{2}{1'}{4}{1}{2}\big]\\
	\cross&\dird{3}{2'}{1}{4}\dird{1'}{4}{3}{2}\bigg\}
\end{align}
\begin{align}
	\frac{\mathrm{d}}{\mathrm{d}\Lambda}\zvortex{1}{2}{1}{2}{1'}{2'}{1}{2}=\frac{1}{2\pi}&\sum_{\alpha_{3^\noprime}\alpha_{3'}}\sum_{\alpha_{4^\noprime}\alpha_{4'}}\int\mathrm{d}\omega_3\int\mathrm{d}\omega_4\left(G^\Lambda_{\alpha_3\alpha_{3'}}(\omega_3)S^\Lambda_{\alpha_4\alpha_{4'}}(\omega_4)+G^\Lambda_{\alpha_4\alpha_{4'}}(\omega_4)S^\Lambda_{\alpha_3\alpha_{3'}}(\omega_3)\right)\nonumber\\\nonumber
	\bigg\{\big[-&\xvortex{1}{2}{3}{4}{1'}{2'}{1}{2}\yvortex{3}{4}{1}{2}{3}{4}{1}{2}-\yvortex{1}{2}{3}{4}{1'}{2'}{1}{2}\xvortex{3}{4}{1}{2}{3}{4}{1}{2}\\\nonumber
	+&\zvortex{1}{2}{3}{4}{1'}{2'}{1}{2}\dvortex{3}{4}{1}{2}{3}{4}{1}{2}+\dvortex{1}{2}{3}{4}{1'}{2'}{1}{2}\zvortex{3}{4}{1}{2}{3}{4}{1}{2}\big]\\\nonumber
	\cross&\dird{1'}{2'}{3}{4}\dird{3}{4}{1}{2}
	\\\nonumber
	-2\sum_{i_3}&\zvortex{1}{4}{1}{3}{1'}{4}{1}{3}\zvortex{3}{2}{4}{2}{3}{2'}{3}{1}\dird{1'}{4}{1}{3}\dird{3}{2}{4}{2}
	\\\nonumber
	+\big[-&\zvortex{1}{4}{1}{3}{1'}{4}{1}{2}\xvortex{3}{2}{2}{4}{3}{2'}{2}{2}- \zvortex{1}{4}{1}{3}{1'}{4}{1}{2}\yvortex{3}{2}{2}{4}{3}{2'}{2}{2}\\\nonumber
	+&\zvortex{1}{4}{1}{3}{1'}{4}{1}{2}\zvortex{3}{2}{2}{4}{3}{2'}{2}{2}+ \zvortex{1}{4}{1}{3}{1'}{4}{1}{2}\dvortex{3}{2}{2}{4}{3}{2'}{2}{2}\big]\\\nonumber
	\cross&\dird{1'}{4}{1}{3}\dird{3}{2'}{2}{4}\\\nonumber
	+\big[-&\xvortex{1}{4}{3}{1}{1'}{4}{1}{1}\zvortex{3}{2}{4}{2}{3}{2'}{1}{2}-\yvortex{1}{4}{3}{1}{1'}{4}{1}{1}\zvortex{3}{2}{4}{2}{3}{2'}{1}{2}\\\nonumber
	+&\zvortex{1}{4}{3}{1}{1'}{4}{1}{1}\zvortex{3}{2}{4}{2}{3}{2'}{1}{2}+\dvortex{1}{4}{3}{1}{1'}{4}{1}{1}\zvortex{3}{2}{4}{2}{3}{2'}{1}{2}\big]\\\nonumber
	\cross&\dird{1'}{4}{3}{1}\dird{3}{2'}{4}{2}
	\\\nonumber
	+\big[&\xvortex{3}{2}{1}{4}{3}{2'}{1}{2}\yvortex{1}{4}{3}{2}{1'}{4}{1}{2}+\yvortex{3}{2}{1}{4}{3}{2'}{1}{2}\xvortex{1}{4}{3}{2}{1'}{4}{1}{2}\\\nonumber
	+&\zvortex{3}{2}{1}{4}{3}{2'}{1}{2}\dvortex{1}{4}{3}{2}{1'}{4}{1}{2}+\dvortex{3}{2}{1}{4}{3}{2'}{1}{2}\zvortex{1}{4}{3}{2}{1'}{4}{1}{2}\big]\\
	\cross&\dird{3}{2'}{1}{4}\dird{1'}{4}{3}{2}\bigg\}
\end{align}
\begin{align}
	\frac{\mathrm{d}}{\mathrm{d}\Lambda}\dvortex{1}{2}{1}{2}{1'}{2'}{1}{2}=\frac{1}{2\pi}&\sum_{\alpha_{3^\noprime}\alpha_{3'}}\sum_{\alpha_{4^\noprime}\alpha_{4'}}\int\mathrm{d}\omega_3\int\mathrm{d}\omega_4\left(G^\Lambda_{\alpha_3\alpha_{3'}}(\omega_3)S^\Lambda_{\alpha_4\alpha_{4'}}(\omega_4)+G^\Lambda_{\alpha_4\alpha_{4'}}(\omega_4)S^\Lambda_{\alpha_3\alpha_{3'}}(\omega_3)\right)\nonumber\\\nonumber
	\bigg\{\big[&\xvortex{1}{2}{3}{4}{1'}{2'}{1}{2}\xvortex{3}{4}{1}{2}{3}{4}{1}{2}+\yvortex{1}{2}{3}{4}{1'}{2'}{1}{2}\yvortex{3}{4}{1}{2}{3}{4}{1}{2}\\\nonumber
	+&\zvortex{1}{2}{3}{4}{1'}{2'}{1}{2}\zvortex{3}{4}{1}{2}{3}{4}{1}{2}+\dvortex{1}{2}{3}{4}{1'}{2'}{1}{2}\dvortex{3}{4}{1}{2}{3}{4}{1}{2}\big]\\\nonumber
	\cross&\dird{1'}{2'}{3}{4}\dird{3}{4}{1}{2}
	\\\nonumber
	-2\sum_{i_3}&\dvortex{1}{4}{1}{3}{1'}{4}{1}{3}\dvortex{3}{2}{4}{2}{3}{2'}{3}{1}\dird{1'}{4}{1}{3}\dird{3}{2}{4}{2}
	\\\nonumber
	\big[&\dvortex{1}{4}{1}{3}{1'}{4}{1}{2}\xvortex{3}{2}{2}{4}{3}{2'}{2}{2}+\dvortex{1}{4}{1}{3}{1'}{4}{1}{2}\yvortex{3}{2}{2}{4}{3}{2'}{2}{2}\\\nonumber
	+&\dvortex{1}{4}{1}{3}{1'}{4}{1}{2}\zvortex{3}{2}{2}{4}{3}{2'}{2}{2}+\dvortex{1}{4}{1}{3}{1'}{4}{1}{2}\dvortex{3}{2}{2}{4}{3}{2'}{2}{2}\big]\\\nonumber
	\cross&\dird{1'}{4}{1}{3}\dird{3}{2'}{2}{4}
	\\\nonumber
	+\big[&\xvortex{1}{4}{3}{1}{1'}{4}{1}{1}\dvortex{3}{2}{4}{2}{3}{2'}{1}{2}+\yvortex{1}{4}{3}{1}{1'}{4}{1}{1}\dvortex{3}{2}{4}{2}{3}{2'}{1}{2}\\\nonumber
	+&\zvortex{1}{4}{3}{1}{1'}{4}{1}{1}\dvortex{3}{2}{4}{2}{3}{2'}{1}{2}+\dvortex{1}{4}{3}{1}{1'}{4}{1}{1}\dvortex{3}{2}{4}{2}{3}{2'}{1}{2}\big]\\\nonumber
	\cross&\dird{1'}{4}{3}{1}\dird{3}{2'}{4}{2}
	\\\nonumber
	+\big[&\xvortex{3}{2}{1}{4}{3}{2'}{1}{2}\xvortex{1}{4}{3}{2}{1'}{4}{1}{2}+\yvortex{3}{2}{1}{4}{3}{2'}{1}{2}\yvortex{1}{4}{3}{2}{1'}{4}{1}{2}\\\nonumber
	+&\zvortex{3}{2}{1}{4}{3}{2'}{1}{2}\zvortex{1}{4}{3}{2}{1'}{4}{1}{2}+\dvortex{3}{2}{1}{4}{3}{2'}{1}{2}\dvortex{1}{4}{3}{2}{1'}{4}{1}{2}\big]\\
	\cross&\dird{3}{2'}{1}{4}\dird{1'}{4}{3}{2}\bigg\}
\end{align}
\newpage

\section{Keldsyh vs. Matsubara}
\renewcommand{\arraystretch}{1.5}
\begin{table}
	\centering
	\begin{tabular*}{0.75\linewidth}{@{\extracolsep{\fill}}l|cc}
		\hline
		\hline
        Step & Matsubara & Keldysh\\
        \hline
        Vertex storage &&\\
        \hline
        Classes & 3 ($s/t/u$) & 3 ($s/t/u$)\\
        Frequencies&$3n_{\omega(\Gamma)}$ ($\omega_c,\nu_c,\nu_c'$)&3$n_{\omega(\Gamma)}$($\omega_c,\nu_c,\nu_c'$)\\
        Locations&$n$&$n$\\
        Types & 2($s/d$) or 4($x/y/z/d$) &2($s/d$) or 4($x/y/z/d$)\\
        Contour/Keldysh indices& 1 & 15 \\
        Data type& \texttt{double}&\texttt{complex<double>}\\
		\hline
        Selfenergy storage & & \\
        \hline
        Frequencies&$n_{\omega(\Sigma)}$&$n_{\omega(\Sigma)}$\\
        Contour/Keldysh indices&1&3\\
        Data type& \texttt{double}&\texttt{complex<double>}\\
		\hline
        Initialization &$\Gamma_{ij}=J_{ij}/4$&$\Gamma^{abcd}_{ij}=J_{ij}/4\begin{cases}\frac{1}{2}&(a=c\land b\neq d)\\&\lor(a=c\land b\neq d)\\0&\text{otherwise}\end{cases}$\\
        \hline
        Internal summations &$\int\mathrm{d}\omega$& $\sum_{\alpha^{(\prime)}}\int\mathrm{d}\omega$ with $\alpha\in0,\dots,8$\\
        \hline
        Propagator calculation & $G(\omega)=1/(\omega-\Sigma(\omega))$&$G(\omega)=(G^{-1}_0-\Sigma)^{-1}$\\
        \hline
        \hline
	\end{tabular*}  
	\caption{Differences and similarities between the Matsubara and Keldysh formalism implementations. These include necessary data structures to store the selfenergy and the vertex for the Keldysh and the Matsubara implementations in the asymptotic frequency parametrization, the initialization, internal summations and the propagator calculation.} 
\label{tab:MatsubaraKeldysh}
\end{table}
To further deepen the comparison between the Matsubara and the Keldysh formalism, we show certain key points for the implementation, especially the data structures in Tab.~\ref{tab:MatsubaraKeldysh}. Due to the Keldysh indices, the Keldysh implementation necessitates more terms and also the \texttt{complex<double>} data type. This also changes the initialization and the internal summations, which in the Keldysh formalism feature an additional summation over the propagator Keldysh indices of which there are only nine due to causality ($G^{11}=0=S^{11}$). Also the propagator calculation requires a matrix inversion instead of a division, since the Keldysh indices make the propagator matrix valued. Further, the symmetries valid throughout the flow are different, which we already presented in the main text and can be compared to e.g.~\cite{DissBuessen,PFFRGReview}. This means, that in principle, a Matsubara PFFRG code can be expanded to the Keldysh formalism if one respects the modified symmetry relations and the differences in internal summation and data structure.
\section{Vertex comparison RPA - FRG}\label{app:comparion}
To show the equivalence of the FRG and the RPA results we show the comparision of the results on vertex level. In the single-channel flow we only get two independent Keldysh indices, which can be chosen to be e.g. $(0,0,0,1)\leftrightarrow\alpha=1$ and $(0,0,1,1)\leftrightarrow\alpha=3$. All other components can be related to these by symmetry operations as e.g. complex conjugation $(1\leftrightarrow 2)$. We show these independent components $\alpha=1,3$ in Fig.~\ref{fig:RPAComparison}. There one can see, that the FRG results with $N_\text{FRG}=50$ are nearly identical to the RPA results with a much higher resolution of $N_\text{RPA}=300$. The resolution for the FRG was chosen to be similar to the FRG flows which were performed for the full Heisenberg model to highlight potential difficulties of the FRG. The notable difference between the FRG and RPA results is a wiggle around $\omega=1$ for all independent vertices, albeit being most noticeable for $i=0,\alpha=1$. This is a result of the remeshing procedure, as we refine our frequency mesh during the flow to accurately capture the essential features at each energy scale. This would have not been necessary for the RPA, as the RPA flow is independent for each frequency, thus allowing us to perform the RPA calculations for each frequency. However, this is not true for the full FRG. Nevertheless, these effects can be minimized by increasing the number of frequencies $N_\text{FRG}$ or optimizing the mesh parameters and seem to play a minor role in the calculation of all our observables. Due to this, we can say, that the FRG accurately reproduces the single step RPA results, highlighting the robustness of the flow and the solving method for the differential equations. 

\begin{figure}[t]
	\centering
	\includegraphics[width=.85\linewidth]{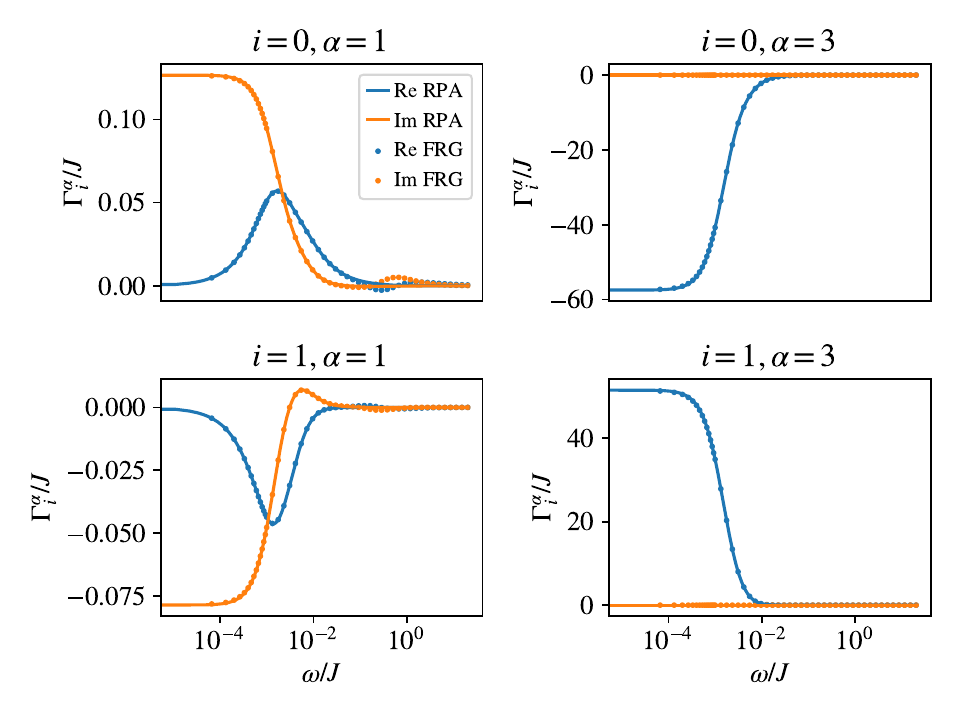}
	\caption{Independent vertex components in the large $S$-limit at $\Lambda=0.002 J\ll T=0.2J$. The FRG data results from a single channel FRG flow with $N_\text{FRG}=50$ while the RPA data is computed in a single step for the given cutoff with a higher resolution of $N_\text{RPA}=300$. The results are nearly identical on vertex level and only a remeshing artifact around $\omega=J$ is visible for the FRG data. 
	}
	\label{fig:RPAComparison}
\end{figure}

\section{Dimer results at lower temperature}\label{app:dimer}
At lower temperatures, the dimer exhibits a kink in the flow, which signals a divergence in the susceptibility and thus ordering tendencies. Since the FRG flow is only defined in the paramagnetic phase, the flow would need to be stopped at this point. The first time, a small increase after the saturation in the nearest neighbor correlator is observable is at $T=0.1J$, albeit being almost non-detectable. This is shown in Fig.~\ref{fig:DimerFlowDiffT}. There, we also show the flow at $T=0.01J$, where the breakdown of the flow is visible much clearer. The breakdown of the flow is also further supported by the double Gaussian fit, as described in the main text (cf. Eq.~\eqref{eq:doubleExponential}). The extracted broadenings -- shown in Fig.~\ref{fig:DimerFitResults} -- which should be correlated to the temperature, feature a step around $T=0.2$. This further indicates the onset of an ordered phase, where the FRG is not valid anymore.
Apart from that, we also show the peak positions from the physical peak in Fig.~\ref{fig:DimerFitResults}, where one can see that the physical value of $\omega/J=1$ is only crossed but the $T\to0$ value does not saturate to. Instead, we see a monotonous increase when decreasing the temperature. This further highlights the findings in the main text, where we should not use the method below $T=0.1J$.

\begin{figure*}[t]
	
	\includegraphics[width=.49\linewidth]{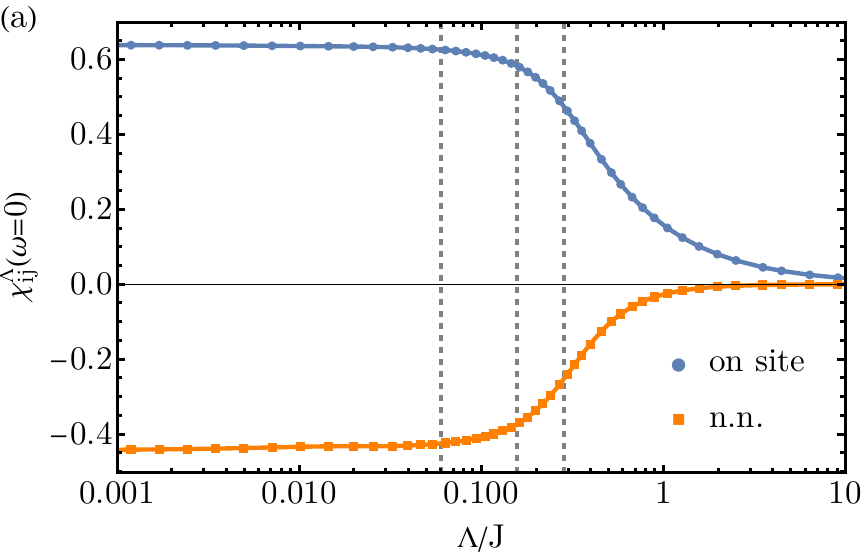}
	\includegraphics[width=.49\linewidth]{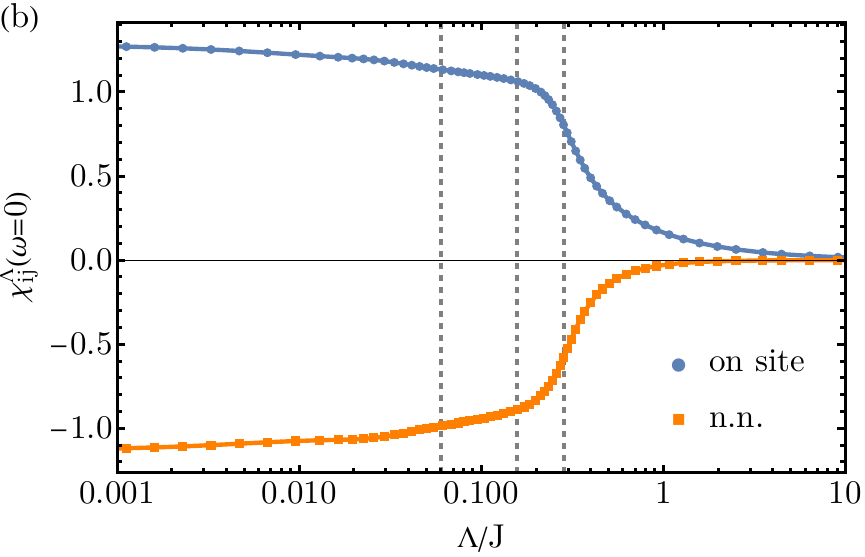}
	\caption{FRG Flow for the AFM Dimer at (a) $T=0.1J$ and (b) $T=0.01J$. The flow exhibits unphysical saturation points, which hint at a breakdown of the flow. Due to this, these results should only be trusted up to the cutoff of the divergence.
	}
	\label{fig:DimerFlowDiffT}
\end{figure*}

\begin{figure*}[t!]
	\includegraphics[width=.49\linewidth]{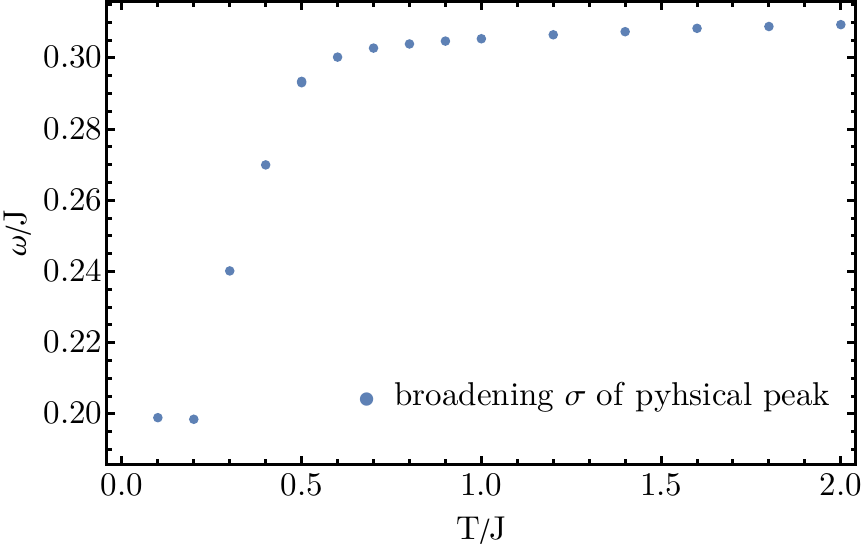}
	\includegraphics[width=.48\linewidth]{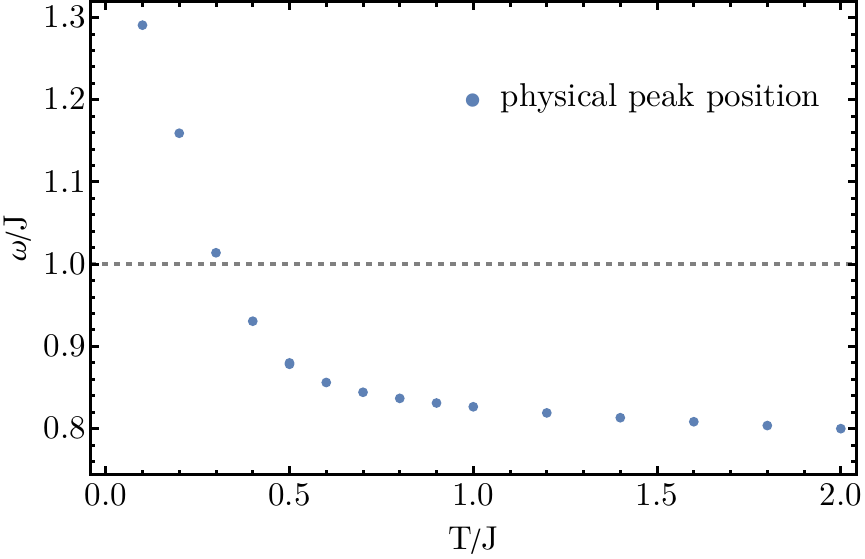}
	\caption{Fit parameters for the AFM dimers at different temperatures. Given are only the values of the physical peak. The peak position shifts with temperature, while the correct value of $\omega/J$ is only passed. The broadening of the peak seems to be fixed around $0.2J$, which could be due to the slight bump, which first emerges at $T=0.1J$
	}
	\label{fig:DimerFitResults}
\end{figure*}

\newpage

\section{Extended symmetry table}\label{app:symmetry}
\begin{table*}[h]          
	\centering
	\begin{tabular*}{0.75\linewidth}{@{\extracolsep{\fill}}lc}
		\hline
		\hline
		\rule[-7pt]{0pt}{23pt}
		$\Sigma^{\alpha'|\alpha}(\omega)=-\zeta(\bm{\alpha})[\Sigma^{\alpha|\alpha'}(\omega)]^*$ & (CC)\\
		\rule[-7pt]{0pt}{23pt}
		$\Sigma^{\alpha'|\alpha}(\omega)=-\Sigma^{\alpha|\alpha'}(-\omega)$ & (PH)\\
		\hline
		\hline
		\rule[-7pt]{0pt}{23pt} $\Gamma_{\mu\nu\,i_1i_2}^{s,\,\alpha'_1\alpha'_2|\alpha^\noprime_1\alpha^\noprime_2}(s,\nu_s,\nu'_s)=\zeta(\bm{\alpha})[\Gamma_{\nu\mu\,i_2i_1}^{s,\,\alpha'_2\alpha'_1|\alpha^\noprime_2\alpha^\noprime_1}(-s,\nu_s,\nu'_s)]^*$ & (X $\circ$ CC $\circ$ PH1 $\circ$ PH2)\\
		\rule[-7pt]{0pt}{23pt} $\Gamma_{\mu\nu\,i_1i_2}^{s,\,\alpha'_1\alpha'_2|\alpha^\noprime_1\alpha^\noprime_2}(s,\nu_s,\nu'_s)=-\xi(\mu)\phantom{[}\Gamma_{\nu\mu\,i_2i_1}^{u,\,\alpha'_2\alpha^\noprime_1|\alpha^\noprime_2\alpha'_1}(s,-\nu_s,\nu'_s)\phantom{]^*}$ & (PH2 $\circ$ X)\\     
		\rule[-7pt]{0pt}{23pt} $\Gamma_{\mu\nu\,i_1i_2}^{s,\,\alpha'_1\alpha'_2|\alpha^\noprime_1\alpha^\noprime_2}(s,\nu_s,\nu'_s)=-\xi(\nu)\phantom{[}\Gamma_{\mu\nu\,i_1i_2}^{u,\,\alpha'_1\alpha^\noprime_2|\alpha^\noprime_1\alpha'_2}(s,\nu_s,-\nu'_s)\phantom{]^*}$ & (PH2) \\
		\rule[-7pt]{0pt}{23pt} $\Gamma_{\mu\nu\,i_1i_2}^{s,\,\alpha'_1\alpha'_2|\alpha^\noprime_1\alpha^\noprime_2}(s,\nu_s,\nu'_s)=\zeta(\bm{\alpha})[\Gamma_{\nu\mu\,i_2i_1}^{s,\,\alpha^\noprime_2\alpha^\noprime_1|\alpha'_2\alpha'_1}(s,\nu'_s,\nu_s)]^*\phantom{-}$ & (X $\circ$ CC) \\
		\hline
		\rule[-7pt]{0pt}{23pt} $\Gamma_{\mu\nu\,i_1i_2}^{t,\,\alpha'_1\alpha'_2|\alpha^\noprime_1\alpha^\noprime_2}(t,\nu_t,\nu'_t)=\zeta(\bm{\alpha})[\Gamma_{\mu\nu\,i_2i_1}^{t,\,\alpha^\noprime_1\alpha^\noprime_2|\alpha'_1\alpha'_2}(-t,\nu_t,\nu'_t)]^*$ & (CC)\\
		\rule[-7pt]{0pt}{23pt} $\Gamma_{\mu\nu\,i_1i_2}^{t,\,\alpha'_1\alpha'_2|\alpha^\noprime_1\alpha^\noprime_2}(t,\nu_t,\nu'_t)=-\xi(\mu)\phantom{[}\Gamma_{\mu\nu\,i_2i_1}^{t,\,\alpha^\noprime_1\alpha'_2|\alpha'_1\alpha^\noprime_2}(t,-\nu_t,\nu'_t)\phantom{]^*}$ & (PH1)\\     
		\rule[-7pt]{0pt}{23pt} $\Gamma_{\mu\nu\,i_1i_2}^{t,\,\alpha'_1\alpha'_2|\alpha^\noprime_1\alpha^\noprime_2}(t,\nu_t,\nu'_t)=-\xi(\nu)\phantom{[}\Gamma_{\mu\nu\,i_1i_2}^{t,\,\alpha'_1\alpha^\noprime_2|\alpha^\noprime_1\alpha'_2}(t,\nu_t,-\nu'_t)\phantom{]^*}$ & (PH2) \\
		\rule[-7pt]{0pt}{23pt} $\Gamma_{\mu\nu\,i_1i_2}^{t,\,\alpha'_1\alpha'_2|\alpha^\noprime_1\alpha^\noprime_2}(t,\nu_t,\nu'_t)=\zeta(\bm{\alpha})[\Gamma_{\nu\mu\,i_2i_1}^{t,\,\alpha^\noprime_2\alpha^\noprime_1|\alpha'_2\alpha'_1}(t,\nu'_t,\nu_t)]^*\phantom{-}$ & (X $\circ$ CC) \\
		\hline
		\rule[-7pt]{0pt}{23pt} $\Gamma_{\mu\nu\,i_1i_2}^{u,\,\alpha'_1\alpha'_2|\alpha^\noprime_1\alpha^\noprime_2}(u,\nu_u,\nu'_u)=\zeta(\bm{\alpha})[\Gamma_{\nu\mu\,i_2i_1}^{u,\,\alpha^\noprime_2\alpha^\noprime_1|\alpha'_2\alpha'_1}(-u,\nu_u,\nu'_u)]^*$ & (X $\circ$ CC)\\
		\rule[-7pt]{0pt}{23pt} $\Gamma_{\mu\nu\,i_1i_2}^{u,\,\alpha'_1\alpha'_2|\alpha^\noprime_1\alpha^\noprime_2}(u,\nu_u,\nu'_u)=-\xi(\mu)\phantom{[}\Gamma_{\nu\mu\,i_2i_1}^{s,\,\alpha^\noprime_2\alpha'_1|\alpha'_2\alpha^\noprime_1}(u,-\nu_u,\nu'_u)\phantom{]^*}$ & (X $\circ$ PH2)\\     
		\rule[-7pt]{0pt}{23pt} $\Gamma_{\mu\nu\,i_1i_2}^{u,\,\alpha'_1\alpha'_2|\alpha^\noprime_1\alpha^\noprime_2}(u,\nu_u,\nu'_u)=-\xi(\nu)\phantom{[}\Gamma_{\mu\nu\,i_1i_2}^{s,\,\alpha'_1\alpha^\noprime_2|\alpha^\noprime_1\alpha'_2}(u,\nu_u,-\nu'_u)\phantom{]^*}$ & (PH2) \\
		\rule[-7pt]{0pt}{23pt} $\Gamma_{\mu\nu\,i_1i_2}^{u,\,\alpha'_1\alpha'_2|\alpha^\noprime_1\alpha^\noprime_2}(u,\nu_u,\nu'_u)=\zeta(\bm{\alpha})[\Gamma_{\mu\nu\,i_2i_1}^{u,\,\alpha^\noprime_1\alpha^\noprime_2|\alpha'_1\alpha'_2}(u,\nu'_u,\nu_u)]^*\phantom{-}$ & (CC) \\
		\hline
		\hline
	\end{tabular*}  
	\caption{Symmetry table for the two particle vertices expanded in terms of Pauli matrices [see Eq.~\eqref{eq:vertexDecomp}] in the asymptotic frequency decomposition. The symmetries are denoted for the $s,t$ and $u$ channel and act as described in the main text.}  
	\label{tab:VertexSymmetry}                             
\end{table*}
In the following we will highlight the symmetries of the self-energy $\Sigma$ and the two particle vertex functions $\Gamma$. We will do so in the asymptotic parametrization \cite{AsymptoticParametrization}, as it is used in the implementation. We use the following definition for the pre-factors
\begin{align}
	\zeta(\bm{\alpha})&=(-1)^{\sum_{i}(\alpha'_i+\alpha^\noprime_i)},\\
	\xi(\mu)&=\begin{cases}
		+1 &\text{$\mu=0$}\\
		-1 &\text{$\mu=x,y,z$}
	\end{cases}.
\end{align}
These pre-factors are a consequence of the symmetry under complex conjugation ($\zeta(\bm{\alpha})$) or due to the particle hole symmetries and a resulting reordering of the Pauli matrices ($\xi(\mu)$). 

\section{Susceptibility calculation}\label{app:Susceptibility}
We start our calculations from Eq.~\eqref{eq:susceptibility} by inserting Eq.~\eqref{eq:GFdecomposition} and writing out all necessary quantum numbers.
This leads to the following Green's functions

	\begin{align}
		G^{12|11}_{\alpha\beta|\alpha'\beta',ij}(t,t'|t,t')&=G^{12|11,c}_{\alpha\beta|\alpha'\beta',ij}(t,t'|t,t')+G^{1|1}_{\alpha|\alpha',ii}(t,t)G^{2|1}_{\beta|\beta',jj}(t',t')-G^{1|1}_{\alpha|\beta',ij}(t,t')G^{2|1}_{\beta|\alpha',ji}(t',t)\\
		G^{11|12}_{\alpha\beta|\alpha'\beta',ij}(t,t'|t,t')&=G^{11|12,c}_{\alpha\beta|\alpha'\beta',ij}(t,t'|t,t')+G^{1|1}_{\alpha|\alpha',ii}(t,t)G^{2|1}_{\beta|\beta',jj}(t',t')-G^{1|2}_{\alpha|\beta',ij}(t,t')G^{1|1}_{\beta|\alpha',ji}(t',t)\\
		G^{22|21}_{\alpha\beta|\alpha'\beta',ij}(t,t'|t,t')&=G^{22|21,c}_{\alpha\beta|\alpha'\beta',ij}(t,t'|t,t')+G^{2|2}_{\alpha|\alpha',ii}(t,t)G^{2|1}_{\beta|\beta',jj}(t',t')-G^{2|1}_{\alpha|\beta',ij}(t,t')G^{2|2}_{\beta|\alpha',ji}(t',t)\\
		G^{21|22}_{\alpha\beta|\alpha'\beta',ij}(t,t'|t,t')&=G^{21|22,c}_{\alpha\beta|\alpha'\beta',ij}(t,t'|t,t')+G^{2|2}_{\alpha|\alpha',ii}(t,t)G^{1|2}_{\beta|\beta',jj}(t',t')-G^{2|2}_{\alpha|\beta',ij}(t,t')G^{1|2}_{\beta|\alpha',ji}(t',t)
	\end{align}

Since we assume that the Green's functions are diagonal in spin space all terms with $G_{\alpha|\alpha'}$ and $G_{\beta|\beta'}$ vanish. Further we use that for $t\neq t'$ the Green's function $G^{1|1}(t|t')$ vanishes. This assumption breaks for $t=t'$ but since this expression is always inside a Fourier integral and has vanishing weight we can neglect this edge case. The remaining terms are 
\begin{align}
	G^{12|11}_{\alpha\beta|\alpha'\beta',ij}(t,t'|t,t')&=G^{12|11,c}_{\alpha\beta|\alpha'\beta',ij}(t,t'|t,t')\\
	G^{11|12}_{\alpha\beta|\alpha'\beta',ij}(t,t'|t,t')&=G^{11|12,c}_{\alpha\beta|\alpha'\beta',ij}(t,t'|t,t')\\
	G^{22|21}_{\alpha\beta|\alpha'\beta',ij}(t,t'|t,t')&=G^{22|21,c}_{\alpha\beta|\alpha'\beta',ij}(t,t'|t,t')\nonumber\\
	&-G^{2|1}_{\alpha|\beta',ij}(t,t')G^{2|2}_{\beta|\alpha',ji}(t',t)\\
	G^{21|22}_{\alpha\beta|\alpha'\beta',ij}(t,t'|t,t')&=G^{21|22,c}_{\alpha\beta|\alpha'\beta',ij}(t,t'|t,t')\nonumber\\
	&-G^{2|2}_{\alpha|\beta',ij}(t,t')G^{1|2}_{\beta|\alpha',ji}(t',t)
\end{align}
Since the FRG implementation is formulated in frequency space we have to perform a Fourier transformation according to eqs.~(\ref{eq:FourierToOmega}) and (\ref{eq:FourierToTime}).
To further decompose the two particle connected Green's function we use the tree expansion \cite{NegeleOrland}
\begin{align}
	G(12|1'2')=-\!\!\!\sumint_{33'44'}\!\!\!G(1|3')G(3|1')\Gamma(3'4'|34)G(2|4')G(4|2')
\end{align}
where one then has to insert the decomposition of the vertex (cf. Eq.~\eqref{eq:decomposition}) and evaluate the sums over the spin indices. We also dropped the indices $c$ denoting connected Green's functions from this point as all following Green's functions will be single particle functions which cannot be disconnected.

For the Heisenberg case ($\Gamma_{\mu\rho}=\Gamma_s\delta_{\mu\rho}$) we then get

	\begin{align}\nonumber
		\chi^\text{Ret}_{ij}(\nu)=\frac{\mathrm{i}}{8\pi}\,&\int\mathrm{d}\omega\left( G^\text{K}(\omega)G^\text{Ret}(\omega+\nu)+G^\text{Av}(\omega)G^\text{K}(\omega+\nu)\right)\delta_{ij}\\
		+\frac{\mathrm{i}}{16\pi^2}&\sum_{\alpha_{3^\noprime}\alpha_{3'}\alpha_{4^\noprime}\alpha_{4'}}\int\mathrm{d}\omega\int\mathrm{d}\omega'\bigg\{\,2\Gamma_{s\,ij}^{\alpha_{3'}\alpha_{4'}\alpha_3\alpha_4}(\omega+\nu,\omega'|\omega,\omega'+\nu)\nonumber	\\\nonumber
		&+\,\Gamma_{s\,ii}^{\alpha_{3'}\alpha_{4'}\alpha_4\alpha_3}(\omega+\nu,\omega'|\omega'+\nu,\omega)\delta_{ij}-\Gamma_{d\,ii}^{\alpha_{3'}\alpha_{4'}\alpha_4\alpha_3}(\omega+\nu,\omega'|\omega'+\nu,\omega)\delta_{ij}\bigg\}\\\nonumber
		\bigg\{&G^{1|\alpha_{3'}}(\omega+\nu)G^{\alpha_3|1}(\omega)G^{2|\alpha_{4'}}(\omega')G^{\alpha_4|1}(\omega'+\nu)\\\nonumber
		+&G^{1|\alpha_{3'}}(\omega+\nu)G^{\alpha_3|1}(\omega)G^{1|\alpha_{4'}}(\omega')G^{\alpha_4|2}(\omega'+\nu)\\\nonumber
		+&G^{2|\alpha_{3'}}(\omega+\nu)G^{\alpha_3|2}(\omega)G^{2|\alpha_{4'}}(\omega')G^{\alpha_4|1}(\omega'+\nu)\\
		+&G^{2|\alpha_{3'}}(\omega+\nu)G^{\alpha_3|2}(\omega)G^{1|\alpha_{4'}}(\omega')G^{\alpha_4|2}(\omega'+\nu)\bigg\}
	\end{align}

This means that we have to perform a numerical integral in two dimensions which makes this computation more complicated as opposed to the normal evaluation of the flow equations. Since however, the results of the spin susceptibility are final, the integrals do not need to be as accurate as within the flow, where errors get scaled during the flow. Note that in the large $S$-limit only the sum term and the propagator term survive \cite{LargeS} while the other contributions are negligible.

For the second case investigated in the paper, which is the Kitaev model the susceptibility has to be evaluated for each spin component separately due to the lack of an $SU(2)$ symmetry. Calculating the spin sums from the vertex decomposition results in

	\begin{align}
		\sum_{\alpha\alpha'\beta\beta'}\sigma^x_{\alpha'\alpha}\sigma^x_{\beta'\beta}\Gamma^{\alpha_{3'}\alpha_{4'}|\alpha_3\alpha_4}_{\alpha\beta|\alpha'\beta',ij}=	\bigg\{&\,4\Gamma_{xx\,ij}^{\alpha_{3'}\alpha_{4'}\alpha_3\alpha_4}(\omega+\nu,\omega'|\omega,\omega'+\nu)\\\nonumber
		-&2\Gamma_{xx\,ii}^{\alpha_{3'}\alpha_{4'}\alpha_4\alpha_3}(\omega+\nu,\omega'|\omega'+\nu,\omega)\delta_{ij}
		+2\Gamma_{yy\,ii}^{\alpha_{3'}\alpha_{4'}\alpha_4\alpha_3}(\omega+\nu,\omega'|\omega'+\nu,\omega)\delta_{ij}\\\nonumber
		+&2\Gamma_{zz\,ii}^{\alpha_{3'}\alpha_{4'}\alpha_4\alpha_3}(\omega+\nu,\omega'|\omega'+\nu,\omega)\delta_{ij}
		-2\Gamma_{d\,ii}^{\alpha_{3'}\alpha_{4'}\alpha_4\alpha_3}(\omega+\nu,\omega'|\omega'+\nu,\omega)\delta_{ij}\bigg\}\\\nonumber
		&\delta(\nu+\nu').
	\end{align}
	\begin{align}
		\sum_{\alpha\alpha'\beta\beta'}\sigma^y_{\alpha'\alpha}\sigma^y_{\beta'\beta}\Gamma^{\alpha_{3'}\alpha_{4'}|\alpha_3\alpha_4}_{\alpha\beta|\alpha'\beta',ij}=	\bigg\{&\,4\Gamma_{yy\,ij}^{\alpha_{3'}\alpha_{4'}\alpha_3\alpha_4}(\omega+\nu,\omega'|\omega,\omega'+\nu)\\\nonumber
		+&2\Gamma_{xx\,ii}^{\alpha_{3'}\alpha_{4'}\alpha_4\alpha_3}(\omega+\nu,\omega'|\omega'+\nu,\omega)\delta_{ij}
		-2\Gamma_{yy\,ii}^{\alpha_{3'}\alpha_{4'}\alpha_4\alpha_3}(\omega+\nu,\omega'|\omega'+\nu,\omega)\delta_{ij}\\\nonumber
		+&2\Gamma_{zz\,ii}^{\alpha_{3'}\alpha_{4'}\alpha_4\alpha_3}(\omega+\nu,\omega'|\omega'+\nu,\omega)\delta_{ij}
		-2\Gamma_{d\,ii}^{\alpha_{3'}\alpha_{4'}\alpha_4\alpha_3}(\omega+\nu,\omega'|\omega'+\nu,\omega)\delta_{ij}\bigg\}\\\nonumber
		&\delta(\nu+\nu').
	\end{align}
	\begin{align}
		\sum_{\alpha\alpha'\beta\beta'}\sigma^z_{\alpha'\alpha}\sigma^z_{\beta'\beta}\Gamma^{\alpha_{3'}\alpha_{4'}|\alpha_3\alpha_4}_{\alpha\beta|\alpha'\beta',ij}=	\bigg\{&\,4\Gamma_{zz\,ij}^{\alpha_{3'}\alpha_{4'}\alpha_3\alpha_4}(\omega+\nu,\omega'|\omega,\omega'+\nu)\\\nonumber
		+&2\Gamma_{xx\,ii}^{\alpha_{3'}\alpha_{4'}\alpha_4\alpha_3}(\omega+\nu,\omega'|\omega'+\nu,\omega)\delta_{ij}
		+2\Gamma_{yy\,ii}^{\alpha_{3'}\alpha_{4'}\alpha_4\alpha_3}(\omega+\nu,\omega'|\omega'+\nu,\omega)\delta_{ij}\\\nonumber
		-&2\Gamma_{zz\,ii}^{\alpha_{3'}\alpha_{4'}\alpha_4\alpha_3}(\omega+\nu,\omega'|\omega'+\nu,\omega)\delta_{ij}
		-2\Gamma_{d\,ii}^{\alpha_{3'}\alpha_{4'}\alpha_4\alpha_3}(\omega+\nu,\omega'|\omega'+\nu,\omega)\delta_{ij}\bigg\}\\\nonumber
		&\delta(\nu+\nu').
	\end{align}

\section{Susceptibility calculation using the Lehmann representation}\label{app:Lehmann}
In this section we are highlighting the calculation for the exact dimer susceptibilities using the Lehmann representation. For this we start by generally calculating the retarded response. For this we use
\begin{align}\label{eq:retSus}
	\chi^{\mu\nu,\text{Ret}}_{ij}(t-t')&=\theta(t-t')\left(\chi^{\mu\nu,>}_{ij}(t-t')-\chi^{\mu\nu,<}_{ij}(t-t')\right)
\end{align}
The greater component is given by 
\begin{align}\nonumber
	\chi^{\mu\nu,\,>}_{ij}&(t,t')=\mathrm{i}\expval{S^\mu_i(t)S^\nu_j(t')}\\&=\frac{\mathrm{i}}{Z}\sum_n\mel{n}{S^\mu_i(t)S^\nu_j(t')\mathrm{e}^{-\beta H}}{n}\\
	&=\frac{\mathrm{i}}{Z}\sum_n\mathrm{e}^{-\beta E_n}\mel{n}{\mathrm{e}^{\mathrm{i}H(t-t')}S^\mu_i\mathrm{e}^{-\mathrm{i}H(t-t')}S^\nu_j}{n}\nonumber\\
	&=\frac{\mathrm{i}}{Z}\sum_{nn'}\mel{n}{S^\mu_i}{n'}\!\!\mel{n'}{S^\nu_j}{n}\mathrm{e}^{-\beta E_n}\mathrm{e}^{\mathrm{i}(E_n-E_{n'})(t-t')}\nonumber
\end{align}
Here we used the Heisenberg picture for the spin operators and inserted an identity operation $\sum\ket{n'}\bra{n'}$. Note that we did not need to introduce time ordering since the spin operators are located on different branches, and thus contour ordering already fixes the order.
Since this expression is only dependent on the time difference $t-t'$ we can reduce this expression to one time variable $t$. 
The lesser susceptibility is calculated analogously and reads
\begin{align}
	\chi^{\mu\nu,\,<}_{ij}(t)=\frac{\mathrm{i}}{Z}\sum_{nn'}\mel{n}{S^\mu_i}{n'}\!\!\mel{n'}{S^\nu_j}{n}\mathrm{e}^{-\beta E_{n'}}\mathrm{e}^{\mathrm{i}(E_{n}-E_{n'})t}.
\end{align}
In this expression we renamed $n\leftrightarrow n'$ to achieve a similar structure.
Inserting the expressions into Eq.~\eqref{eq:retSus} yields
\begin{align}
	\chi^{\mu\nu,\text{Ret}}_{ij}(t)=\frac{\mathrm{i}}{Z}\theta(t)\sum_{nn'}&\mel{n}{S^\nu_j}{n'}\!\!\mel{n'}{S^\mu_i}{n}\\&\left(\mathrm{e}^{-\beta E_n}-\mathrm{e}^{-\beta E_{n'}}\right)\mathrm{e}^{\mathrm{i}(E_{n'}-E_{n})t}.\nonumber
\end{align}
Due to the $\theta(t)$ we cannot perform a Fourier transformation directly but have to introduce a small imaginary shift $\eta$ to ensure convergence as for the free fermions
\begin{align}
	\chi^{\mu\nu,\text{Ret}}_{ij}(\omega)&=\lim_{\eta\to0}\frac{\mathrm{i}}{Z}\sum_{nn'}\left(\dots\right)\left[\frac{\mathrm{e}^{\mathrm{i}(E_n-E_{n'}+\omega+\mathrm{i}\eta)t}}{\mathrm{i}(\omega+\mathrm{i}\eta+E_n-E_{n'})}\right]^{\infty}_0\\
	&=\frac{-1}{Z}\sum_{nn'}\!\frac{\mel{n}{S^\mu_i}{n'}\!\mel{n'}{S^\nu_j}{n}}{\omega+\mathrm{i}\eta+E_n-E_{n'}}\!\left(\mathrm{e}^{-\beta E_n}-\mathrm{e}^{-\beta E_{n'}}\right)\nonumber
\end{align}
As for the Hamiltonian we choose the Heisenberg dimer with and without pseudo-fermion representation and $\mu=\nu=z$. The eigenstates are then $\ket{\uparrow\uparrow}$, $\ket{\downarrow\downarrow}$, $1/\sqrt{2}(\ket{\uparrow\downarrow}+\ket{\downarrow\uparrow})$ and $1/\sqrt{2}(\ket{\uparrow\downarrow}-\ket{\downarrow\uparrow})$. The partition functions are already given in the main text. Inserting the Hamiltonian then gives

\begin{align}
	\chi^R_{00}&(\omega)=\lim_{\eta\to0}\frac{1}{4Z}\bigg[\frac{-2}{\omega+\mathrm{i}\eta}\left(\mathrm{e}^{-\beta\frac{J}{4}}-\mathrm{e}^{-\beta\frac{J}{4}}\right)\\
	&+\left(\frac{1}{\omega+\mathrm{i}\eta-J}-\frac{1}{\omega+\mathrm{i\eta+J}}\right)\left(\mathrm{e}^{-\beta\frac{J}{4}}-\mathrm{e}^{\beta\frac{3J}{4}}\right)\bigg],\nonumber\\
	\chi^R_{01}&(\omega)=\lim_{\eta\to0}\frac{1}{4Z}\bigg[\frac{-2}{\omega+\mathrm{i}\eta}\left(\mathrm{e}^{-\beta\frac{J}{4}}-\mathrm{e}^{-\beta\frac{J}{4}}\right)\\
	&+\left(\frac{1}{\omega+\mathrm{i}\eta+J}-\frac{1}{\omega+\mathrm{i\eta-J}}\right)\left(\mathrm{e}^{-\beta\frac{J}{4}}-\mathrm{e}^{\beta\frac{3J}{4}}\right)\bigg].\nonumber
\end{align}
For better understanding we can use 
\begin{align}
	\lim_{\eta\to0}\frac{1}{\omega+\mathrm{i}\eta}\rightarrow\mathcal{P}\left(\frac{1}{x}\right)-\mathrm{i}\pi\delta(x)
\end{align}
to see that the susceptibility consists of two peaks with opposite signs at $\omega=J$, which come from the first two terms. The last term is vanishing for all frequencies, as the exponentials cancel each other. However, for $\omega=0$ one has to treat the susceptibility carefully since we formally get a derivative
\begin{align}
	\lim_{E_{n'}\to E_{n^\noprime}}\lim_{\eta\to0}\frac{\mathrm{e}^{-\beta E_n}-\mathrm{e}^{-\beta E_{n'}}}{\mathrm{i}\eta+E_n-E_{n'}}.
\end{align}
Performing this derivative leads to the expressions eqs.~(\ref{eq:SusExact}) in the main text.

\section{Errors due to pseudo-fermion decomposition}
Following up on the exact susceptibility calculation in the Lehmann representation, we can use this framework to analyze finite size spin chains. This is necessary due to the fact that the pseudo-fermion decomposition introduces unphysical states into the Hamiltonian, which are either doubly or unoccupied sites and therefore in the context of the magnetic Heisenberg model missing sites. This means, that e.g. for the spin chain a missing sites cuts off longer ranged correlation and essentially reduces the problem to a finite spin chain. In this section we want to analyze the potential influence this can have on the results and compare to the Keldysh PFFRG results to see whether these effects are present.
For the computation of the exact results we only consider physical (i.e. without the pseudo-fermion decomposition) spin chains of $n$ spins. This means that for $n=2$ we recover the dimer, which will be important for the discussion. 

Before introducing the results we first have to discuss how we are Fourier transforming the finite spin chain results, since due to the bi-local nature of the susceptibility we should have two independent momenta $q$ and $q'$. For the infinite lattice we assumed symmetries like the translation symmetry but also rotation or other lattice symmetries to allow us to write the two point correlators $\chi_{ij}(\omega)\equiv\chi(\omega,r_i,r_j)=\chi(\omega,r_\text{rel})$ where $r_\text{rel}=r_i-r_j$ is the relative distance between the two points $i$ and $j$, while the absolute position is not relevant for the susceptibility. We can use this to show that the Fourier transform of the susceptibility is given by
\begin{align}\nonumber
	\sum_{ij}\mathrm{e}^{\mathrm{i} q r_i}\mathrm{e}^{\mathrm{i} q' r_j}\chi_{ij}(\omega)&=\sum_{ij}\mathrm{e}^{\frac{\mathrm{i}}{2} (q-q') (r_i-r_j)}\mathrm{e}^{\frac{\mathrm{i}}{2} (q+q') (r_i+r_j)}\chi_{ij}(\omega)\\&=\sum_{r_\text{rel}}\mathrm{e}^{\frac{\mathrm{i}}{2} (q-q') r_\text{rel}}\chi(\omega,r_\text{rel})\sum_{r_\text{abs}}\mathrm{e}^{\mathrm{i} (q+q') r_\text{abs}}=\sum_{r_\text{rel}}\mathrm{e}^{\mathrm{i} q r_\text{rel}}\chi(\omega,r_\text{rel})
	\end{align}

    \begin{figure}[t]
	\centering
	\includegraphics[width=.99\linewidth]{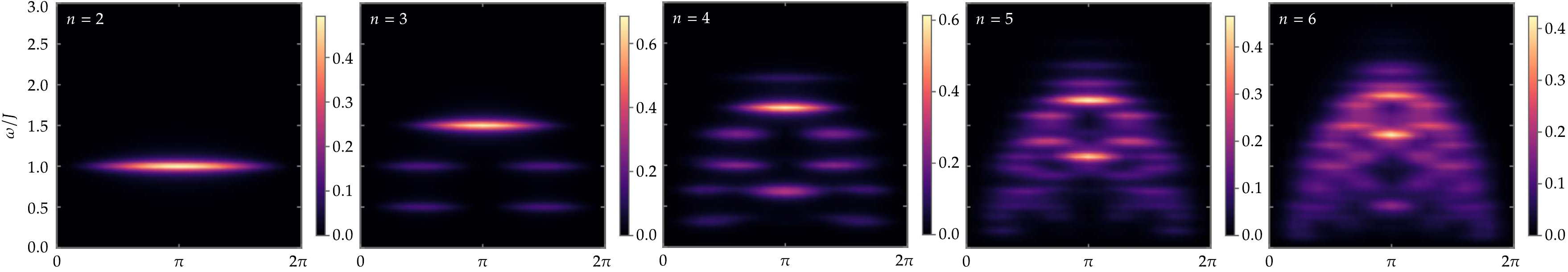}
	\caption{Exact results for the susceptibility $\chi^\text{ret}(\omega,q)$ calculated using the Lehmann representation with $\beta=0.1$ and a finite broadening factor $\eta=0.05$. This can be compared to the Keldysh FRG results from Figure \ref{fig:SpinChainFull} with $J_2=0$. One can see that if we assume the finite spin chains to be perturb the full results, that the broad features of the dimer ($n=2$) at $\omega=J$ and trimer ($n=3$) at $\omega=1.5J$ influence the energy scale and therefore result in the quantitatively wrong frequency scale. Further, the inclusion of the finite spin chains might be the reason for the missing branch to $q=0$, which is also absent in the $n=2$ and $n=3$ results and only become noticeable for $n\geq4$.
	}
	\label{fig:SpinChainExact}
\end{figure}

Here we used that the summation over the relative location results in $\delta(q+q')$ and we therefore have $q'=-q$. To being able to compare to the Keldysh PFFRG results, we therefore evaluate the exact results on the spin chain at $q=-q'$. This might especially seem to be a quiet hard approximation for the dimer with $n=2$ but if we assume the PFFRG to introduce missing sites during the flow, this will be exactly the behavior the PFFRG will naturally have and be therefore a truthful comparison. The results for $n=2,\dots6$ are shown in Figure \ref{fig:SpinChainExact}.
There one can see the expected behavior for the dimer, which is just a broad excitation at $\omega=J$ centered around $q=\pi$ modulated by $\cos(q)$. Paired with the $n=3$ results, this might explain the broad continuum for $J<\omega<1.5J$ in the spin chain results of Figure \ref{fig:SpinChainFull} with $J_2=0$. This definitely is also the reason for the quantitatively incorrect frequency scale, since any emerging frequency scale the results might have are fully influenced by the fixed excitations at $\omega=J$ and $\omega=1.5J$. Therefore we can say, that the pseudo-fermion decomposition results in the inclusion of finite length spin chains into the flow. Additionally this might explain the missing branch to $q=0$, since those are also not formed in the $n=2,3$ results and only slowly start to form for larger $n$. However, since we definitely see features of longer chains, which is the clear maximum at $\omega=0$ for $q=\pi$, this reasoning can only be applied to a certain extent, relying on the suppression of the $q\approx0$ modes. Nevertheless, these problems could very likely be fixed by introducing the pseudo Majorana decomposition instead of the pseudo fermions.

\section{Comparison to VMC}
Since one of the main ways of benchmarking our method in the main text is comparing to the VMC results of F. Ferrari et.~al.~\cite{BeccaSpinChain,VMCSquareLattice,Ferrari-2019} we aim to make the comparison as equal as possible. To achieve the dynamical spin structure factors using VMC one has to convolute Gaussian signals 
\begin{align}\label{eq:Gaussian}
    g(x)=\frac{1}{\Delta\sqrt{2\pi}}\mathrm{e}^{-\frac{(x-x_0)^2}{2\Delta^2}}
\end{align}
with the delta-peak like signals one gets from VMC. In their papers they use a Gaussian with $\Delta=0.1J_1$ for the spin chain and the square lattice and $\Delta=0.02J_1$ for the triangular lattice. This allows them to have sharp branches, which makes sense, as they model the system at zero temperature. In our case however, we are at $T>0$ and therefore we expect additional broadening. As a comparison, we show the dynamical spin structure factor for the spin chain at $J_2=0.7J_1$ with different $\Delta$ compared to our Keldysh PFFRG results at Figure \ref{fig:VMCSmearing}. First of all one notices, that the sharp features smear out significantly and with large broadenings start to resemble the results from the Keldysh PFFRG at finite temperature. Due to this, we chose to use $\Delta=0.2J_1$ in the main text, which still might be a comparably small broadening as opposed to our temperature induced broadening, however, it makes the trends visible without totally obstructing the VMC results as for $\Delta=0.5J_1$. Additionally we chose the case of $J_2=0.7J_1$ due to the large visible gap in the VMC data. This gap is not visible in the Keldysh PFFRG possibly due to the pseudo-fermion constraint violations, as discussed in the main text. Still, this is only a crude approximation, as finite temperature spin structure factors for this model are still missing and thus we cannot compare our results. Nevertheless, one can see that by just increasing the broadening on the $T=0$ results the gap shrinks. We expect this gap to further diminish for $T>0$, thus getting closer to the Keldysh PFFRG results.

\begin{figure}[t]
	\centering
	\includegraphics[width=.99\linewidth]{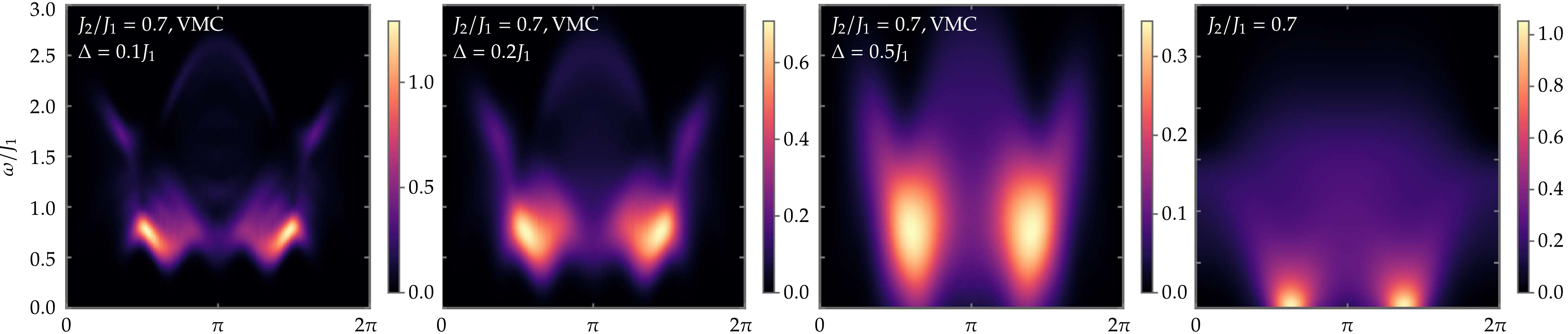}
	\caption{Comparison of the VMC data of Ferrari et. al. \cite{BeccaSpinChain} convoluted with different Gaussian broadenings $\Delta$ according to eq.~(\ref{eq:Gaussian}). The smallest $\Delta$ is the same as the one chosen in their publication, while the larger broadenings are introduced to mimic the effects of temperature effects, which are definitely present in the Keldysh results.
	}
	\label{fig:VMCSmearing}
\end{figure}

\section{Triangular Lattice ground states}
In the main text we analyzed the triangular lattice exemplarily for $J_2/J_1=0.125$ which is inside the spin liquid phase and also for $J_2/J_1=0.5$ which is outside of this spin liquid region and therefore is ordering \cite{Ferrari-2019}. In Fig.~\ref{fig:TriangularGroundstate} we show the real part of the $\omega=0$ suceptibility, which is akin to Matsubara PFFRG a measure for the static susceptibility and therefore the ground state. We could have similarly chosen the static spin structure factor, which looks identical beyond showing a stronger broadening. In these ground state structure factors we can see that at $J_2=0$ we have the typical ordering at the $K$-point, which broadens out until beyond $J_2/J_1=0.08$ the spin liquid phase is formed. This can be seen quiet well at $J_2/J_1=0.07$, where the spectral weight is not solely peaked at the $K$-point but distributed along their connecting lines. For $J_2/J_1=0.125$, the weight is fully distributed, which again supports our claim of being able to describe the spin liquid from the paramagnetic phase. Lastly we show the ground state for $J_2=0.5J_1$, which has shifted the weight to the $M$-points and again indicates ordering. 

Apart from the ground states we also calculated the dynamical spin structure factors shown in Fig.~\ref{fig:TriangularRemaining} for the couplings $J_2=0$ and $J_2/J_1=0.07$, which were not already shown in the main text. The same trend, discussed above is also visible there. However, one would also expect the $M$-point excitiation to be gapped below the phase boundary, which is not discernible in the Keldysh PFFRG.

\begin{figure}[t]
	\centering
	\includegraphics[width=.99\linewidth]{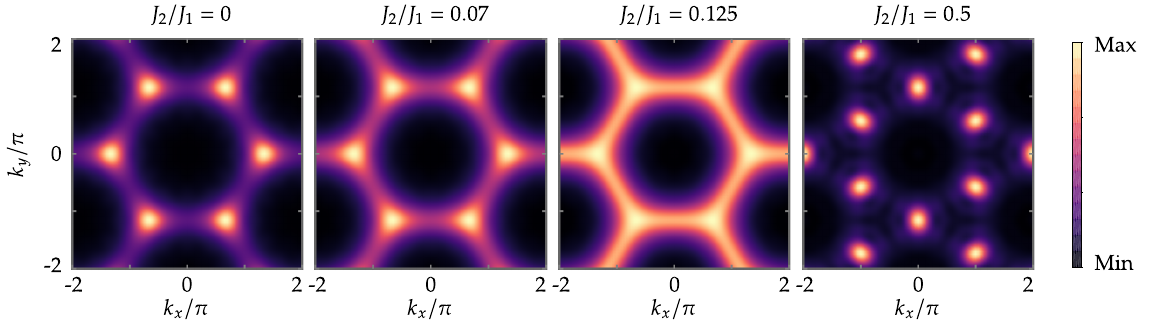}
	\caption{Real part of the $\omega=0$ susceptibility, which is a measure for the static structure factor as used by standard PFFRG\cite{PFFRGReview}. For $J_2/J_1<0.08$ the system is inside the $K$-ordering phase, while for $J_2/J_1=0.125$ we are in the spin liquid phase. $J_2/J_1=0.5$ is then again ordering.
	}
	\label{fig:TriangularGroundstate}
\end{figure}

\begin{figure}[t]
	\centering
	\includegraphics[width=.49\linewidth]{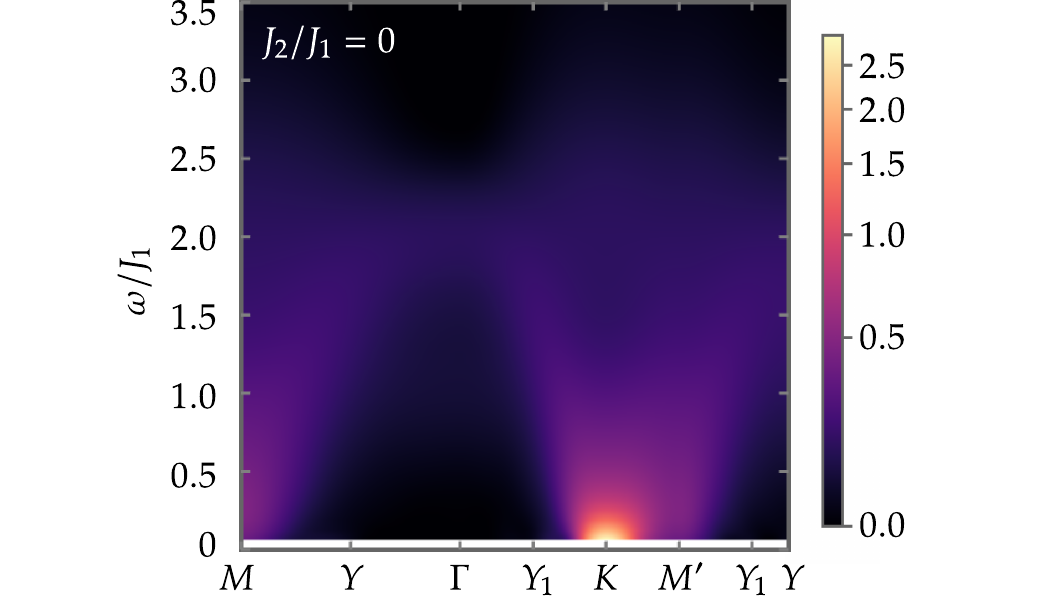}
    \includegraphics[width=.49\linewidth]{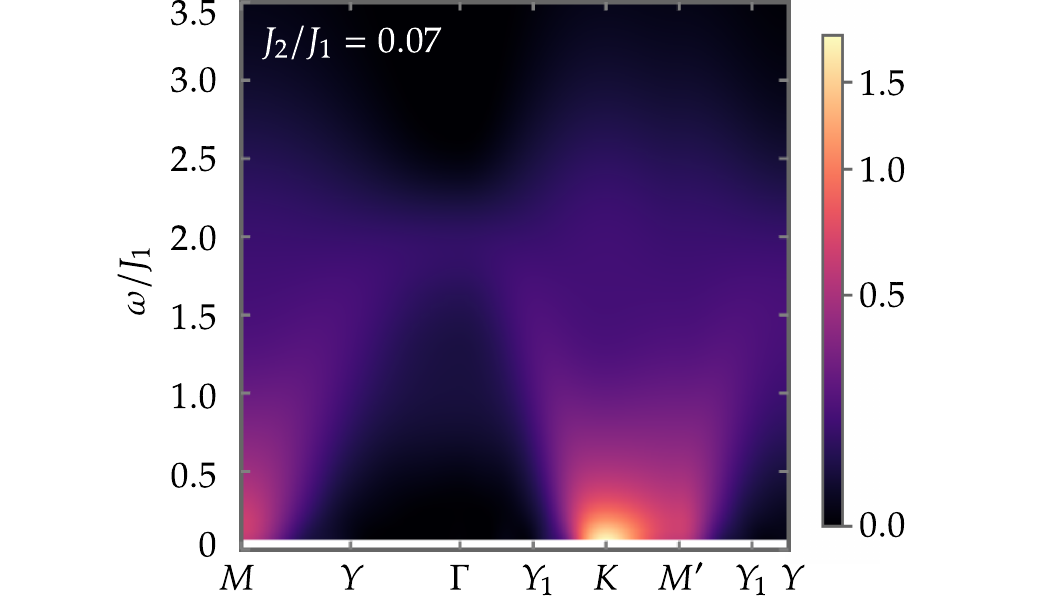}
	\caption{Dynamical spin structure factor for $J_2=0$ and $J_2/J_1=0.07$. The $K$-ordering, which necessitates a Gutzwiller projection in VMC \cite{Ferrari-2019} is directly accessible in Keldysh PFFRG. Approaching the spin liquid phase boundary at around $J_2/J_1\approx0.08$, the weight at the $M$ point increases, as visible for $J_2/J_1\approx0.07$.
	}
	\label{fig:TriangularRemaining}
\end{figure}
\end{widetext}
\end{document}